\documentclass[sigconf]{acmart}
 
\usepackage[T1]{fontenc} 
\usepackage[utf8]{inputenc}
\acmConference[SIGMOD '24]{}{June 03--05,
  2024}{Santiago, Chile}
\usepackage{xcolor}
\usepackage{array}
\usepackage{colortbl}
\if{0}
\documentclass[sigconf, nonacm]{acmart}

\newcommand\vldbdoi{XX.XX/XXX.XX}
\newcommand\vldbpages{XXX-XXX}
\newcommand\vldbvolume{14}
\newcommand\vldbissue{1}
\newcommand\vldbyear{2023}
\newcommand\vldbauthors{\authors}
\newcommand\vldbtitle{\shorttitle} 
\newcommand\vldbavailabilityurl{https://github.com/orm011/seesaw}
\newcommand\vldbpagestyle{plain} 
\fi

\usepackage{etoolbox}
\usepackage{float}
\usepackage{caption}

\usepackage[nodisplayskipstretch]{setspace}

\graphicspath{ {./assets/} }
\usepackage{booktabs} 
\usepackage{markdown}

\usepackage{multirow}
\usepackage{enumitem}

\usepackage[linesnumbered]{algorithm2e}

\usepackage{amsmath}
\usepackage[export]{adjustbox}

\DeclareMathOperator*{\argmax}{argmax}
\DeclareMathOperator*{\argmin}{argmin}

\usepackage{caption}
\usepackage{subcaption}

\makeatletter

\renewcommand{\sectionautorefname}{\S\@gobble}
\renewcommand{\subsectionautorefname}{\S\@gobble}
\renewcommand{\subsubsectionautorefname}{\S\@gobble}
\makeatother

\makeatletter
    \@ifdefinable{\sys}{\def\sys/{SeeSaw}}
\makeatother

\definecolor{forestgreen}{RGB}{34,139,34}

\definecolor{codegreen}{rgb}{0,0.6,0}
\definecolor{codegray}{rgb}{0.5,0.5,0.5}
\definecolor{codepurple}{HTML}{C42043}
\definecolor{backcolour}{HTML}{FFFFFF}
\definecolor{bookColor}{cmyk}{0,0,0,0.90}  
\color{bookColor}

\newcommand\numberstyle[1]{%
    \footnotesize
    \color{codegray}%
    \ttfamily
    \ifnum#1<10 0\fi#1 |%
}

\setlist[itemize]{leftmargin=*,nosep,topsep=0pt}
\setitemize{noitemsep,topsep=0pt,parsep=0pt,partopsep=0pt}

\setlist[enumerate]{leftmargin=*,nosep,topsep=0pt}
\setlist[description]{leftmargin=*,nosep,topsep=0pt}

\newenvironment{reviewerenv}{\itshape}{\ignorespacesafterend}

\newcommand{\revision}[1]{\textcolor{blue}{#1}}
\renewcommand{\revision}[1]{#1}
\newenvironment{revisionenv}{\upshape \color{blue}}{\ignorespacesafterend}

\renewenvironment{revisionenv}{}{\ignorespacesafterend}

\begin{document}

\if{0}
\twocolumn 
\section*{Response to Reviewer Comments}

We thank the reviewers for their constructive comments.  We have addressed all of your comments, including additional experiments on hyperparameters and adding new baselines. We have highlighted changes in the paper \revision{in blue}.

\noindent\rule{\columnwidth}{1pt}

\textbf{R1O1} \begin{reviewerenv} The baselines for the paper are lackluster. Relevance feedback is not a new idea and so more text should be devoted to the connections with its literature and comparing with its existing methods. At the very least, I would a tuned version of a simple baseline like Rocchio's algorithm, which already contains the idea of regularization to initial query vector + movement towards positive examples and away from negative examples. I am less familiar with relevance feedback for image search systems, but I don't believe this is the first system to consider the topic and would expect more complex baselines as well. The idea of comparing only against a system without relevance feedback is insufficient and I'm unsure that people would consider ENS (\cite{Jiang2017-iy}) the state of the art baseline.\end{reviewerenv}

\begin{revisionenv}
    We have added an optimized Rocchio baseline in \autoref{sec:baselines}, with results in \autoref{tab:baselines}.  We tuned the hyperparameters for Rocchio like we did for \sys/, including the commonly used $\gamma=0$\cite{irbook}), as shown in the table below. We used the best setting, highlighted in gray.
    
\begin{table}[ht!]
\caption*{Tuning Rocchio update}
\label{tab:hyperparam_rocchio_update}
\begin{tabular}{rr|rrrr|r}
\toprule
$\beta$ & $\gamma$ & BDD & COCO & LVIS & ObjNet & Avg. \\
\midrule
0.50 & 0.00 & \bfseries 0.77 & 0.96 & 0.72 & 0.68 & 0.78 \\
\hline
\rowcolor{lightgray}
0.50 & 0.25 & 0.75 & \bfseries 0.97 & \bfseries 0.74 & \bfseries 0.70 & \bfseries 0.79 \\
\hline
0.50 & 0.50 & 0.75 & 0.95 & 0.74 & 0.66 & 0.78 \\
0.75 & 0.00 & 0.73 & 0.96 & 0.70 & 0.67 & 0.76 \\
0.75 & 0.25 & 0.76 & 0.96 & 0.73 & 0.68 & 0.78 \\
0.75 & 0.50 & 0.76 & 0.96 & 0.73 & 0.66 & 0.78 \\
\bottomrule
\end{tabular}
\end{table}

With respect to, ENS \cite{Jiang2017-iy} we found that it is the most recent work from a major conference that models  a similar search problem, albeit in an abstract vector setting. It is not a paper from the information retrieval community, so doesn't use the formulation of relevance feedback, instead drawing on terminology and baselines from the active learning community.
    
We observe that other recent work in major conferences, such as \cite{Coleman2022-ag}, has also independently found \cite{Jiang2017-iy} to be a state-of-the-art baseline in a retrieval setting. 
\end{revisionenv}

\textbf{R1O2} \begin{reviewerenv}It's a little bit hard to tell the relative importance of the logistic regression loss and the regularization term just from equation (2). It would be nice to understand given feedback how much the query is changing at each iteration.\end{reviewerenv}

\begin{revisionenv}
The regularization term greatly affects the results.  To help make this dependence clearer, we have also added ``few-shot CLIP'' to  in \autoref{tab:baselines}, in addition to Rocchio's algorithm. Both Rocchio and \sys/ include some type of regularization, which is why both perform better. Few-shot CLIP sometimes achieves AP below that of zero-shot CLIP.

\end{revisionenv} 

\textbf{R1O3} \begin{reviewerenv}
    Can the authors explain how often the query is reformulated for Seesaw.
\end{reviewerenv}

\begin{revisionenv}
For all the experiments in the paper, the query (the query vector) is updated after every image. 
\end{revisionenv}

\textbf{R1O4} \begin{reviewerenv}
The authors should help clarify the computational time spent running the various algorithms in SeeSaw. In particular, the authors discuss many ways in which label propagation is too slow to use in the system. The results should show the related improvements and how this compares to the slower, more complex algorithms in question.
\end{reviewerenv}

\begin{revisionenv}
    We added a section on latency and scalability in \autoref{sec:scalability}, including a latency comparison of SeeSaw and the label propagation version (which uses full propagation rather than a regularization term) in \autoref{tab:latency}, which shows propagation latency reaches multiple seconds for the larger dataset. The problem would be worse with even larger, but common, datasets.
\end{revisionenv}

\textbf{R1O5} \begin{reviewerenv}
The multi-scale technique seems good to explain and makes sense, but I figured this result would be known. Is this unknown for CLIP? I know many object detectors are multi-scale and would guess intuitively that people know object detection for small objects in clip might require multiple scales due to its training process as being from the main caption of an image.
\end{reviewerenv}
\begin{revisionenv}
While it is true that people have used the idea of multiscale methods (sometimes called ``pyramid methods'') in image processing and retrieval, we are not aware of people using it in a CLIP-based image retrieval systems before. Further, as we note in the introduction, while multi-scale is a conceptually simple idea, because it substantially increases the number of vectors we have to index, it demands a very lightweight search algorithm.

We have added a paragraph in the related work section regarding multi-scale representation.

    



\end{revisionenv}

\textbf{R2O1} \begin{reviewerenv}
Unclear hyper-parameter tuning and guaranteed improvements. The additional terms for clip alignment and database alignment have their specific hyper-parameters. It remains unclear how these hyper-parameters should be tuned in the absence of user feedback. I could imagine a spectrum of techniques, but please clarify the deployment process for a new data collection.

\end{reviewerenv}

\begin{revisionenv}
    We found the overall results are stable within a wide range of hyper-parameters values. To clarify this in the paper, we added \autoref{sec:hyperparams}, which includes \autoref{tab:hyperparam_multi_reg}, showing how the overall AP results for all datasets remain high even as we vary all 3 hyperparameters by an order of magnitude each.
    
    Therefore, regarding your question of how hyperparameters would be tuned in the absence of feedback: we believe it is reasonable to reuse the same parameters we used for the datasets we have, as we found one setting of hyperparameters worked well for all of the datasets we used in our experiments. We also state this in the new section.
\end{revisionenv}

\begin{reviewerenv}    
Section \autoref{sec:dbalign} leverages an existing label propagation technique [44], but the introduced matrix expressions seem to have inconsistent dimensions. First, Equation (4) introduced $t(y) @ (D-W) @ y$, further describes $W$ as adjacency matrix and $D$ as diagonal matrix - for squared matrices $D$ and $W$, the expression would be fine, but the paper defines $W$ as a $k$-by-$N$ matrix, which renders the expression invalid due to mismatching dimensions. Second, the same conflict then propagates to expressions like $M_D = t(X_D) @ (D-W) @ X_D$

\end{reviewerenv}

\begin{revisionenv}
    In \autoref{sec:dbalign}, both $D$ and $W$ are  $N\times N$ matrices, so there is no dimension mismatch.  In the original text, we meant $W$ is sparse (because it is a k-nearest neighbor adjacency matrix) so there are only $k \times N$ non-zero elements within that $N\times N$ matrix. This sparsity matters in terms of computational complexity, it is not the same as the matrix dimensions.  We have clarified the text.

\end{revisionenv}

\begin{reviewerenv}
Furthermore, the paper mentions multiple times the aim of providing accuracy improvements with guarantees of no quality regressions. While I appreciate the empirical analysis, please clarify if a hard guarantee can be given upfront. Specifically, the AP would not be very indicative for the long tail of infrequently occurring queries, but would be dominated by the majority classes, right?
\end{reviewerenv}

\begin{revisionenv}
    Our claim is only empirical. We now used the word empirical in the abstract to clarify that early. \autoref{fig:quantiles} backs this empirical robustness claim.
    
    As you correctly point out, a simple average would be dominated by majority classes. This is our motivation for reporting both the CDF of AP in \autoref{fig:quantiles}, as well as the mean value of AP in \autoref{tab:breakdown} aggregated across all queries, but also aggregated only for the long tail alone, which would be hidden in the global average.  We note that our method performs substantially better on tail queries where CLIP does poorly (and perform identical to CLIP in situations where CLIP does well!) 

\end{revisionenv}

\textbf{R2O2} \begin{reviewerenv}
Partially incomplete related work. The paper contains a brief section of related work, and also discusses related work throughout the paper. However, the paper should make more effort to put the presented approach into the context of existing (data management) research. I would recommend to structure the related work by sub-areas, and briefly survey these areas with one-paragraph per area. Relevant areas not covered yet include query relaxation techniques, active learning for data cleaning, entity resolution, and schema matching, as well as data/query provenance such as why-not provenance.
\end{reviewerenv}

\begin{revisionenv}
    We have now structured  \autoref{sec:related} as you suggsested, and include a paragraph discussion for each of these sub-areas mentioned above.
\end{revisionenv}

\textbf{R2O3} \begin{reviewerenv} 
Paper presentation lacks structure (paper organization, missing conclusion). The paper lacks proper structure, which creates an unnecessary burden on the reader. I would recommend to make the following improvements:
\begin{itemize}
\item Add the missing section ``Conclusions'' before the references, which should contain a brief summary, actual conclusions you draw from this work (what did we learn, what can be reused, other insights), and a brief statement on future work.
\begin{revisionenv}
    We have added a conclusion  at the end of the paper, sorry for the omision.
\end{revisionenv}
\item Use more structure in terms of paragraph labels, figures and tables, bullet point list, etc. Right now a user faces walls of text (e.g., Section 1, 3.1, 3.2, 4.1, 4.2, 4.3), which makes is hard to go back and read the details of a specific aspect again. For example (see Section 1 contributions), use bold font for the paragraph label and use a proper bullet lists with indentation for the list of contributions.
\begin{revisionenv}Thank you for the suggestion. We have added paragraph titles in bold throughout these sections  to split the text into more manageable chunks. We have fixed the indentation.
\end{revisionenv}
\item Use the right template (e.g., normally section titles are all-caps). 
\begin{revisionenv}
     Thanks for pointing out the template issues, there was a silent incompatibility between the template and an imported package.
\end{revisionenv}

\item Fix the minor presentations issues
\end{itemize}
\end{reviewerenv}

\textbf{R3O1} \begin{reviewerenv}
The paper could benefit significantly from clearer writing and context establishment in various sections.


In Section 1, the mention of the ``training dataset'' within a vector query processing problem lacks context, causing confusion. It would be beneficial to clarify this concept early on instead of abruptly introducing it. \revision{Done.}

In Section 3.1, the term ``ideal query vector'' is introduced without clear explanation, which might confuse readers. \revision{We have clarified this text.}

Section 3.2. ``Because zero-shot approaches are new, much existing work is not really designed in that context.'' Do you mean ``few-shot approaches''? Otherwise, it's confusing.\revision{Thank you, we have removed this confusing statement.}

The narrative style used in Section 4.2 is somewhat disorganized and lacks structure, making it challenging to trace the logical steps of the presented algorithms. Important concepts, such as $M_D$ appear suddenly without adequate explanation. In addition, some ideas, like the use of sigmoid function, gets ``deprecated'' very quickly to pave way for yet another idea. Readers may question their time investment. 

I recommend that the authors begin Section 4.2 by explaining the goal of database alignment at a high level, focusing on the target problem, the challenges, and how they are addressed. The section could then be restructured according to these points. A similar restructuring could also benefit Sections 4.1 and 4.3.
\revision{We have added additional structure and paragraphs headings throughout to clarify the text.}

Lastly, the paper is missing a conclusion section.  
\begin{revisionenv}
    We have now added the missing conclusion section, we apologize for the omission.
\end{revisionenv}
\end{reviewerenv}

\textbf{R3O2} 
\begin{reviewerenv}
The paper does not address system scalability, an important aspect of any image search system. Questions surrounding how sampling might affect accuracy performance, the size of the evaluated dataset, and the necessity of retuning hyperparameters $\lambda, \lambda_c, and \lambda_D$ with the addition of new images, remain unanswered. These points should be discussed, and the use of illustrative figures might be beneficial to clarify these issues.
\end{reviewerenv}
\begin{revisionenv}
    Regarding the effect of increasing dataset size: dataset size has several effects including possible affecting interactive latency and also whether or not the hyperparameters need to be re-tuned for larger data sets.         
    With respect to the effect of increasing the size of the data set on system latency during interactive querying, we have added section \autoref{sec:scalability}  including \autoref{tab:latency}, showing that latency remains below .5 seconds for different dataset sizes, for \sys/. 
          
    Regarding the re-tuning of hyperparameters, we have found that different datasets do not generally require different hyperparemeter settings, as shown in a new experimental section \autoref{sec:hyperparams}. We discuss this further in our response to R2O1.
    
    Regarding sampling: we do not employ any sampling in our experiments. We mentioned sampling once, as an aside, as a method to speed up the computation of the kNN for a database, as well as the related $M_D$ matrix. 
    
    
\end{revisionenv}
\fi

\title{\sys/: Interactive Ad-hoc Search Over Image Databases}

\author{Oscar Moll}
\email{orm@csail.mit.edu}
\affiliation{%
  \institution{MIT CSAIL}
  \city{Cambridge}
  \state{MA}
  \country{USA}
}

\author{Manuel Favela}
\email{mfavela@alum.mit.edu}
\affiliation{%
  \institution{MIT}
  \city{Cambridge}
  \state{MA}
  \country{USA}
}

\author{Sam Madden}
\email{madden@csail.mit.edu}
\affiliation{%
  \institution{MIT CSAIL}
  \country{USA}
}

\author{Vijay Gadepally}
\email{vijayg@ll.mit.edu}
\affiliation{%
  \institution{MIT Lincoln Laboratory}
  \country{USA}
  }

\author{Michael Cafarella}
\email{michjc@csail.mit.edu}
\affiliation{%
  \institution{MIT CSAIL}
  \country{USA}
  }

\if{0}
\renewcommand{\shortauthors}{Moll, et al.}
\fi

\begin{abstract}
As image datasets become ubiquitous, the problem of ad-hoc searches over image data is increasingly important.  
Many high-level data tasks in machine learning, such as constructing datasets for training and testing object detectors, imply finding ad-hoc objects or scenes within large image datasets as a key sub-problem. 
New foundational visual-semantic embeddings trained on massive web datasets such as Contrastive Language-Image Pre-Training (CLIP) can help users start searches on their own data, but we find there is a long tail of queries where these models fall short in practice.
\sys/ is a system for interactive ad-hoc searches on image datasets that integrates state-of-the-art embeddings like CLIP with user feedback in the form of box annotations to help users quickly locate images of interest in their data even in the long tail of harder queries.  
One key challenge for \sys/ is that, in practice, many sensible approaches to incorporating feedback into future results, including state-of-the-art active-learning algorithms, can worsen results compared to introducing no feedback, partly due to CLIP's high-average performance.
\revision{Therefore, \sys/ includes several algorithms that empirically result in larger and also more consistent improvements.}
We compare \sys/'s accuracy to both using CLIP alone and to a state-of-the-art active-learning baseline and find \sys/ consistently helps improve results for users across four datasets and more than a thousand queries. 
\sys/ increases Average Precision (AP) on search tasks by an average of .08 on a wide benchmark (from a base of .72), and by a .27 on a subset of more difficult queries where CLIP alone performs poorly.
\end{abstract}

\maketitle
\setcounter{page}{1}

\if{0}

\pagestyle{\vldbpagestyle}
\begingroup\small\noindent\raggedright\textbf{PVLDB Reference Format:}\\
\vldbauthors. \vldbtitle. PVLDB, \vldbvolume(\vldbissue): \vldbpages, \vldbyear.\\
\href{https://doi.org/\vldbdoie.}{doi:\vldbdoi}
\endgroup
\begingroup
\renewcommand\thefootnote{}\footnote{\noindent
This work is licensed under the Creative Commons BY-NC-ND 4.0 International License. Visit \url{https://creativecommons.org/licenses/by-nc-nd/4.0/} to view a copy of this license. For any use beyond those covered by this license, obtain permission by emailing \href{mailto:info@vldb.org}{info@vldb.org}. Copyright is held by the owner/author(s). Publication rights licensed to the VLDB Endowment. \\
\raggedright Proceedings of the VLDB Endowment, Vol. \vldbvolume, No. \vldbissue\ %
ISSN 2150-8097. \\
\href{https://doi.org/\vldbdoi}{doi:\vldbdoi} \\
}\addtocounter{footnote}{-1}\endgroup

\ifdefempty{\vldbavailabilityurl}{}{
\vspace{.3cm}
\begingroup\small\noindent\raggedright\textbf{PVLDB Artifact Availability:}\\
The source code, data, and/or other artifacts have been made available at \url{\vldbavailabilityurl}.
\endgroup
}
\fi

\section{Introduction}

Increasingly inexpensive cameras and storage make it ever easier to collect image data.
Large quantities of video and images are now captured from dedicated cameras as well as mobile phones, vehicles, and drones.
Nevertheless, the ability of an engineer or team to explore their own image data and discover ad-hoc items of interest lags far behind their ability to collect that data.

For example,
an engineer at an autonomous vehicle company with a large repository of data may wish to find examples of people in wheelchairs to extend an object detector or to find examples of bikes in the snow to test an existing detector. 
Or, an ornithology researcher may want to search their camera archives to know which of all their camera locations seem to show more sightings of a particular type of bird \cite{wildlife_insights,Van_Horn2015-zu,Young2018-xv}.

Whether the goal is to extend the capabilities of an object detector model, to enhance test cases for existing models, or simply to explore a large trove of image data, finding relevant images within a dataset of images or video is a key problem for many users. 
For example, depending on the dataset, wheelchairs or bikes in the snow may be rare, appearing in only one in a thousand images or less. Hence, ad-hoc searches can be challenging without efficient image search tools.

%
%
A well-known high-level approach to the search problem is representing the contents of the database as well as the queries as vectors: a text query becomes a {\em query vector} $q$, and an image in the database becomes an {\em image vector} $x$, and the relevance of an image $x$ to a query $q$ is estimated by its inner product $q\cdot x$, which measures the geometric {\em alignment} of the query and the database element (every vector is unit normed). The most relevant results for a query vector $q$ are the solution to $\argmax_x x \cdot q$, the elements with a maximum inner product with $q$. One advantage of this modeling approach is the availability of scalable and low-latency vector stores. These stores create an index of all vectors $x$ ahead of time and then solve the search at query time without a full scan of the database of $x$, enabling interactive searches for any query.

This vector paradigm relies on mapping queries and images to vectors, and the deep learning-based method to achieve the goal of mapping images and text to vectors today is through {\em cross-modal} or {\em visual-semantic} embeddings~\cite{Frome2013-ue,Faghri2017-kr,Chen2019-iv,Radford2021-rk,Jia2021-np}. The accuracy of this approach depends on the quality of the cross-modal embedding used to capture the important concepts in an image.

State-of-the-art cross-modal embedding models consist of large neural networks trained on massive datasets of corresponding image-text pairs.
A notable example is CLIP \cite{clip_model}, which is a model pre-trained on a crawl of hundreds of millions of web images and their {\tt alt-text} attributes as a low cost proxy for a caption. 

\begin{figure}[ht]
\caption{The solid line plots the cumulative distribution function of Zero-shot CLIP Average Precision (AP) across four evaluation datasets introduced later. The horizontal dashed line marks the fraction and absolute number of queries with $AP < .5$, for each dataset.}
\label{fig:zero_shot_variance}
\includegraphics[width=\columnwidth]	{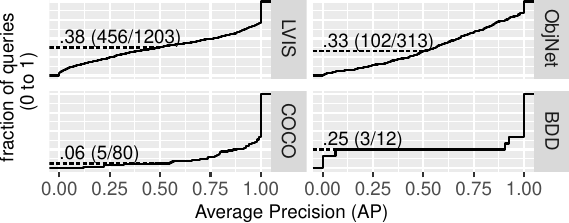}
\end{figure}

\paragraph{{\bf Zero-shot CLIP}} With CLIP, users can often cold start many searches using text alone in new datasets without fine-tuning any model, an approach called {\em zero-shot} learning. Zero-shot CLIP is surprisingly accurate even when used on datasets it was not trained on, including many of our evaluation datasets.  
\paragraph{{\bf Limitations of existing approaches.}}
In the case of CLIP, despite the high average zero-shot quality of the model, the quality of the results on a specific dataset varies substantially between queries.  
For example, using CLIP embeddings on the Berkeley Deep Drive (BDD) dataset \cite{bdd} of street scenes, we can easily find scenes with bicycles.  For wheelchairs, however, which only occur in a handful of scenes, using CLIP alone requires looking through more than 100 images before the first wheelchair is found. \autoref{fig:zero_shot_variance} shows the accuracy distribution in queries from four evaluation datasets, measured as  AP, a common accuracy metric. The step on the right edge for each dataset corresponds to a substantial number of queries with optimal results ($AP=1$). The trailing slope on the left shows a long tail of queries with lower accuracy. The annotations on the dashed line quantify the fraction and amount of queries with $AP < .5$ using zero-shot CLIP, which is large in some of the datasets.   We note that the ``queries'' used in these plots correspond only to labeled categories in these datasets.  From a user's point of view, the plotted distribution is not as relevant as the distribution seen on their queries of interest; i.e., high accuracy on bicycles does not compensate for poor results on wheelchair if that is the query of interest.

\begin{figure}[ht]
     \caption{Two causes of suboptimal search results: alignment deficit (left), and locality deficit (right). Both circles and arrows are unit norm, lying on the unit sphere (dashed circle).}
     \label{fig:problem_diagram}
     \centering     
     \begin{subfigure}[b]{0.4\columnwidth}
         \centering
         \includegraphics[width=\textwidth]{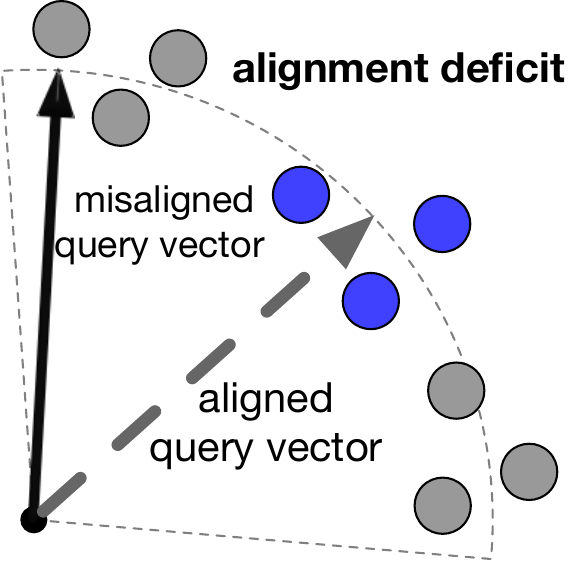}
         \caption{}
         \label{fig:alignment}
     \end{subfigure}
      \unskip\ \vrule\
      \begin{subfigure}[b]{0.4\columnwidth}
         \centering
         \includegraphics[width=\textwidth]{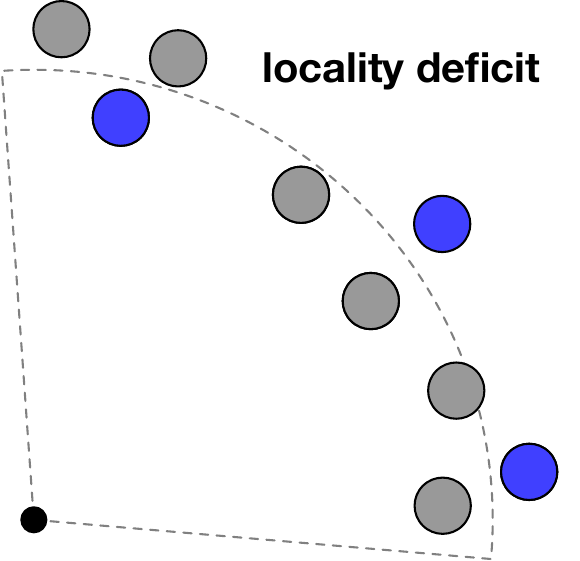}
         \caption{}
         \label{fig:locality_diag}
     \end{subfigure}
     \begin{subfigure}{\columnwidth}
         \centering
         \includegraphics[width=.8\textwidth]{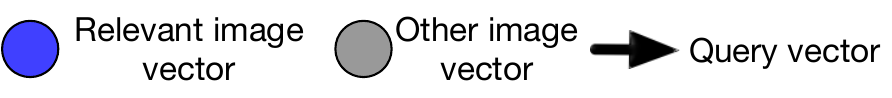}
     \end{subfigure}
\end{figure}

 \paragraph{\bf{Query alignment}} CLIP sometimes provides low-relevance results for two reasons.  First, the CLIP embedding of the string (e.g., ``wheelchair'') may not be close enough  to the relevant image vector embeddings;  we call this a {\em query alignment} deficit. \autoref{fig:alignment} shows a visual representation of lack of query alignment: all the image vectors, represented as circles arranged along an imaginary circular arc (due to image vectors being normalized unit size). The blue colored circles represent vector embeddings of the relevant images (or, as we introduce later, relevant patches within images), in this case images of wheelchairs in BDD. The initial query maps, via the string embedding, to the vector represented by the solid arrow. In the diagram,  this initial query vector aligns with the gray circles at the top more than with the blue circles to the right.  Therefore, these non-relevant results would appear first in a search, ahead of the relevant results, causing difficulties to users.  


\paragraph{\bf{Concept locality}} A second possible problem causing poor results is that embeddings for images of interest (e.g., wheelchairs) may not be clustered tightly together in the database, as shown in (\autoref{fig:locality_diag}). In this diagram, we see that no single query arrow would be able to align well with all three blue image vectors because they are diffused among non-relevant gray vectors. Regardless of what vector a user may conjure up, the results will never be wholly relevant.  We call this situation a {\em concept locality} deficit.  Queries often present both types of deficits, so they benefit from improving either.


\paragraph{\bf{Our solution}: \sys/}  \sys/'s goal is to allow users to search their data leveraging embeddings such as CLIP and helping users improve their results when needed.  Users work in a loop with \sys/ providing feedback in the form of boxes around relevant regions of images.  This process results in better-aligned query vectors, such as the dashed line vector in \autoref{fig:alignment}, thereby improving results. A user interacts with \sys/ following the pseudo-code of \autoref{alg:basic_loop}: a search starts with a text string, which is converted the text into a vector value $q_0$ using an embedding model like CLIP (\autoref{line:clip}). $q_0$ is used as the query vector for a lookup operation into the vector store (\autoref{line:clip}), which locates the most relevant vector $x$ in the store for the query vector, i.e., the one with the largest inner product. The corresponding image is presented to the user, who provides feedback. The query vector is now updated to value $q_1$ via \texttt{query\_align} in \autoref{line:align}, which includes previous feedback. In the next round, $q$ is updated to value $q_2$, and so on. Ideally, results improve on every round of feedback. In reality, each loop consists of a batch of a user specified size.

\begin{algorithm}
\caption{\textbf{ high-level structure of search with user feedback. \sys/ focuses on the logic of} \texttt{query\_align}}
\label{alg:basic_loop}
\SetKwInOut{Input}{input}
\SetKwInOut{Output}{output}
\DontPrintSemicolon
\Input{\ \ text\_query}
feedback\_map $\gets \{\}$\; 
query\_vector $\gets$ CLIP.embed\_string(text\_query)\;\label{line:clip}
\While{True}{
   img\_id $\gets$ vector\_store.lookup(query\_vector)\;\label{line:vectorstore}
   img\_feedback $\gets$ UI.show(img\_id)\;\label{line:ui}
   feedback\_map.update(img\_id, img\_feedback)\;
   query\_vector $\gets$ query\_align(feedback\_map)\;\label{line:align}
}
\end{algorithm}

 \paragraph{{\bf Related work}} The general idea of leveraging user feedback to locate results is known as {\em relevance feedback} in information retrieval~\cite{Salton1990-ks,Jing2004-ex},\cite[Ch.9]{Manning2008-ua}, and as {\em active search} within active learning \cite{Garnett2012-tp}. However, \sys/ must address two challenges not present in prior work: first, we find that basic implementations of \texttt{query\_align} as well as state-of-the-art active search \cite{Jiang2017-iy} can decrease result relevance when the starting point is zero-shot CLIP, even when all approaches start with the same high-quality CLIP embeddings.  Second, practical approaches also need to provide interactive latency for large datasets, meaning the computational work needed on every round should grow sub-linearly with the dataset size, which is not true of state-of-the-art active search approaches.
 
\paragraph{{\bf \sys/ insights}} \sys/ addresses both challenges based on three main insights: 
The first insight is that we should merge user labels with the original CLIP query rather than relying on either alone.
\sys/ accomplishes through its implementation of \texttt{query\_align}  of \autoref{line:align}, within which it searches for a query vector minimizing a custom loss function.  This loss function goes beyond reflecting accuracy on the observed feedback as a standard supervised-learning approach would:  \sys/ additionally encourages similarity between the aligned query and the original embedding of the concept being searched through a novel regularization term, ensuring the CLIP text query is used for both the initial search and also within the minimization problem. We refer to this idea as CLIP alignment.

The second insight is that the process to improve the query should reflect the data distribution of the entire database rather than only that of user feedback: since user feedback data in \sys/ up to iteration $t$ is based \revision{on those vectors in $X_D$} nearest to $q_0$, $q_1$,... and $q_{t-1}$, the observed data is skewed toward this very specific region of the database.  One hypothetical way to address this problem is sampling randomly from the dataset and labeling these nodes, but this approach imposes labeling requirements on the user, and for many hard-to-find objects this approach would find no positive examples. \sys/ cheaply approximates this hypothetical process by adding a second database-dependent regularization term to the loss function. We show this regularization term is conceptually equivalent to synthesizing a new training set where elements are instead picked randomly and uniformly from the vector database, and their unobserved labels are approximated through label-propagation\cite{Zhu2002-hd}. We refer to this  as database alignment or DB alignment. 

\sys/ models both CLIP alignment and DB alignment into a loss function that can be quickly minimized so that \sys/ can produce a better-aligned query vector with little input from the user and with low latency, and so that \sys/ can leverage fast vector stores for search.  Moreover, the amount of work \sys/ does at query time within the loop in \autoref{alg:basic_loop} grows only with the size of the observed dataset,  unlike active learning approaches we evaluate that rely on linear time scoring of the full database after each feedback iteration.

In addition to the above techniques to improve query alignment, \sys/ employs a third insight: a multi-vector, multi-scale representation of images derived from separately embedding patches of different sizes and positions within a single image. This representation is motivated by observing that in complex images the object of interest  may not be the most prominent feature in an image; this is a common cause of accuracy problems for zero-shot CLIP. This technique is conceptually simple and orthogonal to CLIP and DB alignment, but in practice, because this representation multiplies the number of vectors in the database by an order of magnitude,  only search techniques whose latency does not depend linearly on the vector database size can be integrated easily with it.

\vspace{.1in}
\noindent {\bf Contributions.}  In summary, our contributions are:
\begin{enumerate}[labelindent=0pt]
\item We introduce \sys/'s custom {\em query alignment} algorithms for user-in-the-loop image search: CLIP alignment combined with database alignment, which provide high quality results with a fixed and limited amount of user feedback while avoiding any linear time computational costs at query time that could hinder interactivity on larger datasets.
\item We combine the query alignment algorithms with a multi-scale feature representation for images that is possible due to the scalability of the alignment algorithms.
\item We demonstrate with extensive benchmarks across 4 datasets and hundreds of queries, that \sys/ consistently improves retrieval metrics; overall \sys/ improves Average Precision (AP) from .19 to .46 on a subset of more challenging queries.
\item We demonstrate that alternative techniques to implementing relevance feedback can often either reduce search accuracy with respect to the zero-shot CLIP approach, or scale poorly with data, or both.
\end{enumerate}

\section{System}
\label{sec:system}
\begin{figure}[htb]
\caption{\sys/ component diagram. Top: preprocessing steps; Bottom: dataflow during interaction loop.}
\label{fig:system}
\centering
\includegraphics[width=\linewidth]{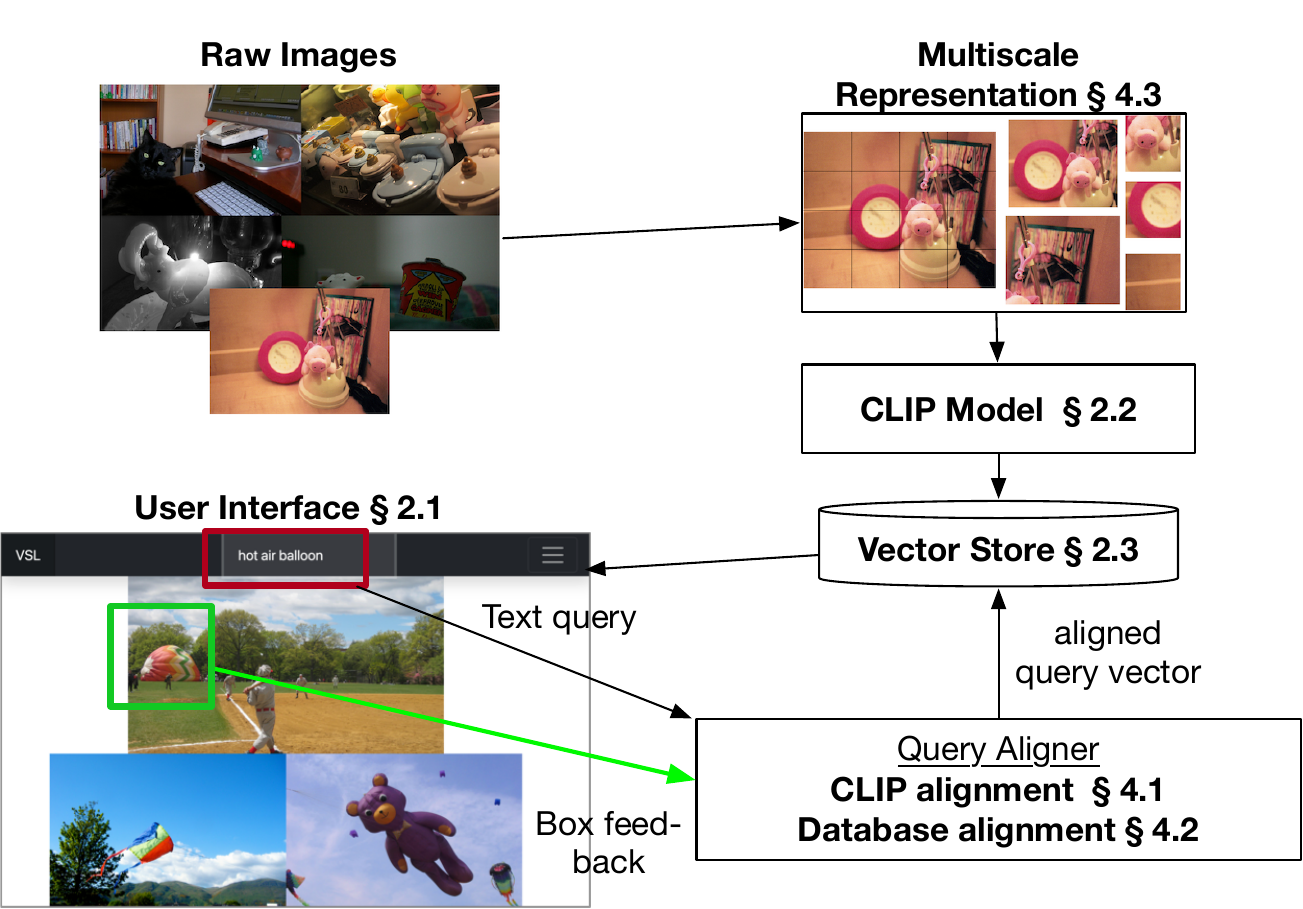}
\end{figure}

\sys/ consists of the following components: 1) a graphical user interface,  2) a pre-trained visual semantic embedding model (CLIP~\cite{Radford2021-rk}), 3) an indexed vector store for max inner product queries (Annoy \cite{annoy}), and 4) a server layer, which we will call the query aligner,  mediating between the other components and implementing the query alignment described above. \autoref{fig:system} shows how data moves between these components.

\subsection{CLIP}
\label{sec:embedding}
\sys/ uses the CLIP \cite{clip_model} pre-trained visual-semantic embedding model for preprocessing and during querying. \sys/ embeds images into vectors using the  visual component of CLIP during preprocessing.  During querying, \sys/ uses the string embedding component of CLIP to translate  string inputs from the user into query vectors.

\subsection{Vector Store} 
\label{sec:vector-store}

After raw image data is processed into vectors, and is indexed by the vector store, the vector store provides a maximum inner product lookup interface with low latency, which is important  because the user waits on results from the system.  
The vector store needs to be accurate, but does not need to be exact: it is acceptable for the result in \autoref{line:vectorstore} of \autoref{alg:basic_loop} to be among the top largest rather than exactly the largest, as even if the exact result were returned, there is already error inherent to the embedding representation as shown in \autoref{fig:locality}

Our implementation uses the Annoy store\cite{annoy}, which offers only {\em approximate} maximum innner product lookup, which is also what most vector stores offer. We saw only a minor drop in accuracy metrics in our benchmarks using Annoy vs an exact but slow scan.

\subsection{User Interface (UI) and Querying}
\autoref{fig:system}, bottom left,  shows a screenshot of the \sys/ UI for the hot air balloon query as one of the components of \sys/.  A user wishing to make a model to detect hot-air balloons can begin the search process with \sys/ through text by typing ``hot-air balloon'' into the search box. The loop of \autoref{alg:basic_loop} runs,  and through the UI, the user provides feedback on the results offered so far (\autoref{line:ui}). This flow of data is diagrammed in \autoref{fig:system}.

\subsection{Preprocessing} 
Before using \sys/, we perform a one-time pre-processing pass over the image data. Pre-processing in \sys/ consists of converting raw image data into semantic feature vectors using a pre-trained visual embedding (CLIP, in our case). For \sys/, the runtime of this preprocessing pipeline depends on four variables: the number of images in the dataset, the pixel sizes of the images in the dataset, the inference cost of the embedding, and the number and type of Graphics Processing Units (GPUs) available. On COCO, a dataset of 120000 images, \sys/ preprocessing in our un-optimized single GPU pipeline takes less than an hour. Because this task is data parallel, the runtime can be reduced to minutes by using more GPUs.  Furthermore, model optimization techniques just as JIT compilation would further reduce this runtime for real applications with larger datasets. 

The vector store, Annoy, takes less than 20 minutes to build the index over the vectors computed above. These costs are incurred once per dataset and are then amortized across all subsequent queries. 

\section{Approach}

The key idea of our approach is to improve query alignment by leveraging user feedback and by enriching that feedback with two other sources of information: the CLIP embedding of the query itself (\autoref{sec:clipalign}) as well as the structured of the unlabeled database (\autoref{sec:dbalign}).  
\sys/ incorporates these different sources of information within a single loss function, which is minimized with respect to an internal query vector parameter on every round to yield the next query vector \sys/ will use internally.
A secondary aspect of our approach is using a multi-vector, multi-scale representation of the data  which we cover in \autoref{sec:multiscale}. 

\subsection{Motivation for Query Alignment}
\label{sub:motivation}

In the introduction we explained how both {\it query alignment} and {\it concept locality} are important for searches to work well (diagrammed in \autoref{fig:problem_diagram}). \sys/ focuses on query alignment, so it is valuable to quantify the potential gains and the limitations of this approach. Because our evaluation datasets have complete labels for different categories, which we will use as evaluation queries, we can understand how far from the ideal any approach is.  In this section, we show a large fraction of labeled categories in the ObjectNet dataset presents a combination of high concept locality, and a lot of the error is due to lower query alignment, and therefore adjusting alignment alone can improve results substantially.

\paragraph{\bf Ideal query vector} For a query such as ``wheelchair'' within ObjectNet, we can measure query alignment deficit by comparing the accuracy of using the embedding of the string ``a wheelchair'' to the accuracy of an {\em ideal query vector} to find wheelchairs within the database, one derived with full knowledge of ObjectNet images and their ground-truth labels. We can compute an ideal query vector by fitting a linear classifier model on the CLIP image embedding vectors $X$, where each embedding is labeled with $y = 1$ or $0$ depending on whether or not the image is labeled as having a wheelchair in it. This linear model is certainly over-fit from a prediction perspective; but, in this case, model fitting is a simple and efficient search method to find out whether there are {\em any} high-accuracy query vectors. 

Using the labels for ``wheelchair'' we can then compute the accuracy of results when using the string-derived query vector or the ideal query vector. We will see shortly that there are queries where even this best-fit vector has low accuracy: a strong indication of low concept locality; and conversely, a common case is when the best-fit vector has high or perfect accuracy, but the string-derived vector has low accuracy, a strong indication of high locality for the concept but low alignment for the initial query, the kind of scenario where \sys/ would work best. 

\begin{figure}[ht]
\caption{ Comparing the Average Precision (AP) of the ideal query vector vs the initial CLIP string embedding for each labeled class on the ObjectNet dataset. Each point in the scatter plot represents one of the 300 categories in ObjectNet.  The median AP  for the ideal queries (see vertical boxplot) is above .9, and more than 25\% reach 1, while the median AP of initial queries (see horizontal boxplot) is around .2; showing that even though concept locality in the embedding is high, the initial query alignment can be relatively poor. Section \autoref{sub:motivation} explains the setup in more detail. }
\label{fig:locality}
\centering
\includegraphics[width=3.3in]{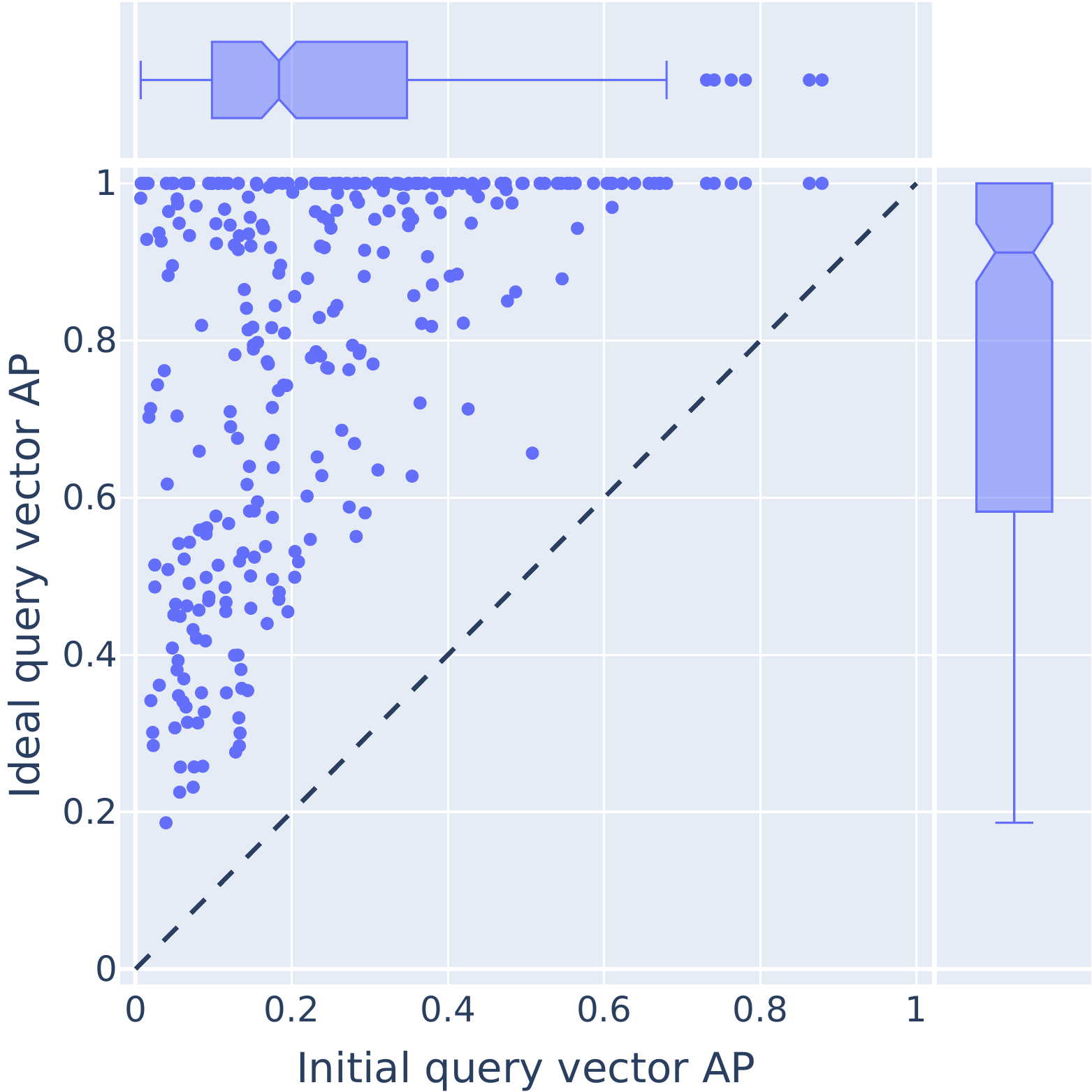}
\end{figure}

After carrying out the above process for 300 different queries, we measure accuracy for each query using  Average Precision scores, and plot the AP of the ideal and initial queries for all 300 ObjectNet labeled categories as the y and x coordinates of scatter-points in \autoref{fig:locality}. 

\paragraph{{\bf Average Precision (AP)}} AP is an accuracy metric common for information retrieval \cite{irbook}, because it rewards earlier relevant results more heavily than simpler metrics such as precision or F-score, and without picking an arbitrary result cutoff. AP values range from 0 to 1. An AP of 1 means perfect accuracy: when all positive results appear before the negative search results. 

The figure shows the median AP for the ideal queries (see vertical boxplot) is above .9, and more than 25\% reach 1, while the median AP of initial queries (see horizontal boxplot) is around .2, which shows that  concept locality in the embedding is high because the ideal vectors perform much better than the string derived query vectors.  Note several best-fit queries score 1, indicating those queries have very high locality and improving alignment is all that is needed. This is not always the case, but even queries with low $y$ coordinate values where locality may be a problem show benefits from improving alignment (hence lie comfortably above the diagonal dashed line).

Note that these numbers do not mean we can easily find ideal vectors in practice, given the lack of labeled data and the very few samples available from feedback to the system, but they show that focusing on alignment makes sense for CLIP embeddings of this dataset. 




\subsection{Motivation for CLIP Alignment Approach}

One natural way to implement query alignment in the context of a text search is to use a CLIP string vector $q_0$ to locate a few examples that we ask the user to label.  This results in a set of examples $(x,y)$ which we can use to learn a new query vector as part of \texttt{query\_align} in \autoref{line:align} in \autoref{alg:basic_loop}. In the simplest approach, we can simply train a standard logistic regression model on the user's labels for results seen so far. After round $t$ of feedback from the user, we pick $q_{t+1}$ as follows:

\begin{equation*}
 q_{t+1} = \argmin_{w} \mathcal{L}_1(\mathbf{w})
\end{equation*}

 \begin{equation}
 \label{eq:lr} 
 \mathcal{L}_1(\mathbf{w}) = 
          \sum_{i=1}^{t}  \text{LogLoss}\left(y_i, \text{sigmoid}(\mathbf{w}^\top \mathbf{x}_i+b ) \right)  +  \lambda |\mathbf{w}|^2 
 \end{equation}

\paragraph{{\bf Few-shot CLIP}}  The summation term is the logistic loss function added over elements with user feedback, the weight vector $\mathbf{w}$ is learned, as is the scalar bias $b$, typical in standard logistic regression and necessary to achieve well-calibrated output probabilities.  In practice we find fitting both $\mathbf{w}$ and $b$ as opposed to forcing $b$ to be 0 substantially reduces the accuracy of the learned $\mathbf{w}$ as a query, so we do not use the $b$ parameter.

The extra term, $\lambda |\mathbf{w}|^2$, where $\lambda$ is a scalar hyperparameter, penalizes large magnitudes in the weight vector and is again a standard penalty applied to the loss function of logistic regression to prevent selecting values of $\mathbf{w}$ that are very large when the data are fully separable~\cite{Murphy2022-au}. In the interactive setting we work in, with very few labeled examples, it is essentially guaranteed that the labeled data will be separable because the number of labeled data items is small compared to the dimension (512) of the CLIP embedding, so this penalty is always necessary. 

This approach based on \autoref{eq:lr} is called the {\em few-shot} CLIP approach, as opposed to the {\em zero-shot} CLIP approach of using $q_0$ derived from a string with no feedback.  Few-shot CLIP has some advantages over zero-shot CLIP because the learned query vector $q_{t+1}$ is now based on actual vectors from the database. Moreover, CLIP embeddings of images show high average few-shot learning accuracy on many datasets: a handful of examples are often enough to train a linear model with high accuracy \cite{Radford2021-rk}.  However, we find the few-shot approach using \autoref{eq:lr} in the way we described is less accurate than the zero-shot CLIP approach, and the accuracy drop is evident empirically on all our datasets, as we will explain and show later in \autoref{tab:breakdown}. The few-shot approach is a baseline in its own right, and we evaluate it together with other baselines in \autoref{sec:baselines}.\\

There are several reasons for the drop in accuracy of the few-shot CLIP approach from the zero-shot CLIP approach:  First, the absolute accuracy of zero-shot CLIP can be high, hence no method can improve substantially on it.  Second, $q_1$ in the few-shot approach is computed from very few vectors from the database, depending on the batch size, unlike the ideal vectors computed in \autoref{fig:locality} which beat the zero-shot CLIP vector but are trained on thousands of samples. In machine learning algorithms small samples lead to larger generalization error. Third, the vectors $X_t,y_t$ from relevance feedback up until round $t$ are not necessarily representative either of the region of the vector space that contains relevant results, nor of the full database, where the learned $q_{t+1}$  will be used as a query.  

We note that on some occasions, when the initial embedding query $q_0$ is of sufficiently poor quality,  the few-shot CLIP approach does improve results beyond what the zero-shot approach offers. However, even in that case, the approaches we introduce offer strong advantages over either approach alone, as we show in our evaluation.






\section{Detailed Approach}
\label{sec:approach}

\sys/ leverages the previous high-level insights into multiple techniques which  we explain in detail in this section. First, we integrate the zero-shot and few-shot approaches into a single combined approach.  We call this approach {\em CLIP alignment}. Second, we guide our query vector improvement process to also account for the structure of the unlabeled vectors in the database, which we call {\em Database alignment}.  Finally, we add a multiscale image representation, where we extract multiple vectors for different patches of the image at preprocessing time to allow us to capture objects that appear in different scales and positions in images.

\subsection{CLIP Alignment}
\label{sec:clipalign}

Consider a search scenario where we first find a single positive example $x_0$ and then a single negative example $x_1$, in that order, using the CLIP string embedding as the initial query vector $q_0$.
There are many possible query vectors that produce the same observed ordering of $x_0,x_1$: for example, clearly $q_0$ produces this order, as does $x_0$ used as a query itself, which is guaranteed to score itself as the maximum possible score of 1. A third possible query vector is found by minimizing a loss function such as logistic loss (similar to \autoref{eq:lr}) which would align somewhat with the positive example and away from the negative example. There are many more possible solutions that produce this same observed ranking.

However, all these solutions will produce potentially different search results in the next round, and these will have different qualities.
How should we select the query vector? A typical approach is cross-validation, where data is repeatedly separated into test and training sets to learn a model that generalizes well on unseen data, but in this context with a handful of points, cross-validation is not feasible.

Instead, we rely on a rule-of-thumb based on the {\em principle of stability} for machine learning algorithms \cite[p. 174]{Shalev-Shwartz2014-kf}. 
The principle states that when choosing between two methods that both fit the data equally well, the method more likely to generalize is the one that changes least when we include or exclude a data point from the training set.
Intuitively, a method that overly relies on any one data point $x$ is also most susceptible to generalization errors due to sample variance.
In our setting, the original query vector $q_0$ of zero-shot CLIP is not influenced by the observed data at all, so there are reasons to prefer that vector as our next query by default.

\paragraph{{\bf CLIP alignment loss term}} In reality, however, our initial query vector $q_0$ will often fall short, as we showed in the $x$-axis of  Figure~\ref{fig:locality}, for example. In many cases we must strike a balance between being unduly influenced by sample noise and incorporating information from feedback.  This observation is the basis of CLIP alignment, which we implement by adding an extra penalty term to the loss function in \autoref{eq:lr}:  

\begin{equation}
\label{eq:clip_alignment}
 \mathcal{L}_2(\mathbf{w}) =  \mathcal{L}_1(\mathbf{w}) + \lambda_{c} \left( 1. - \mathbf{w} \cdot q_0/|\mathbf{w}| \right)
\end{equation}

 The term $(\mathbf{w} \cdot q_0)/|\mathbf{w}|$ is the cosine distance between the parameter $\mathbf{w}$ and the initial query $q_0$ (which is normalized), encouraging the optimal $\mathbf{w}$ to geometrically align with the original CLIP text query $q_0$, in addition to minimizing the previous loss. 
 
 
 In other words, if two possible query vectors $w_1$ and $w_2$ have the same classification loss, the one with the highest cosine similarity to the original query $q_0$ will be favored by this loss function.
  
 $\lambda_c$ is a new hyper-parameter that governs the trade-off between fitting the feedback data and preserving the new weight's similarity to the original. A large $\lambda_{c}$ parameter means we ignore the user labels and a small one means we ignore the initial text query. 

As more user examples come in, the user input is weighted more highly with respect to the CLIP prior.
We show in the evaluation section that the resulting query vectors from adding this additional loss term are more accurate than {\em either} the original $q_0$ (the zero-shot approach) or one learned purely from the data (the few-shot approach), as in \autoref{eq:lr}, and this is the case even when $q_0$ has a poor initial performance.  

\subsection{Database (DB) Alignment}
\label{sec:dbalign}
From a supervised learning point of view, a query minimizing \autoref{eq:clip_alignment} on a large sample is also likely to show a low error over the full database. However, in \sys/ we minimize  \autoref{eq:clip_alignment} over a small sample $x_t$ of examples previously shown to the user, for which we got feedback $y_t$.   Besides their small size, these samples are not a random sample of the database $X_D$ because they were elements with high similarity to previous query vectors $q_{t}, q_{t-1} ..., q_0$. This is an instance of a {\em domain shift} problem, where the target  domain distribution   $X_D$ differs from the training set distribution $X_t,y_t$, even if the mapping $X \to y$ being learned is the same \cite{Kouw2021-vr}.

\sys/ uses this observation as a starting point to further improve query alignment with the real database. At a high level, our key observation is that we can approximate an unbiased and large sample of the data $X_D, y_D$ using the label propagation algorithm defined in \cite{Zhu2002-hd}, which we will explain below,  and then use this new training set consisting of $X_D$ and $\hat{y}_D$ when solving for $\mathbf{w}$ in \autoref{eq:clip_alignment}. 

Note that while the ``propagated'' labels $\hat{y}_D$ are only an estimate of the unobserved labels, the distribution of $X_D$ faithfully reflects the distribution of the database vectors.

We found the propagation step improves the final classifier, but that the propagation algorithm can reduce interactivity, as it must run after every round of feedback to propagate the new labels and  requires iterating over a full k-Nearest Neighbor graph (kNN graph) of the vectors in the database. 

\paragraph{\bf{DB alignment loss term}} Hence, the version of database alignment we use in \sys/ takes the propagation approach only as a conceptual starting point but achieves the effect by adding a second alignment term to the loss function \autoref{eq:clip_alignment}, producing \autoref{eq:db_alignment}. 

 In this updated loss function, the matrix $M_D$ in the expression  $\mathbf{w}^T M_D \mathbf{w}$, which we will explain in detail shortly,  depends on the vectors in the database $X_D$, and on their kNN graph, but not on the query $q_0$, so it can be computed once per dataset ahead of time.  $M_D$'s size is $512 \times 512$, and is constant with respect to the size of the database: its size is only a function of the CLIP embedding dimension of 512, not of dataset size.

\begin{equation}
\label{eq:db_alignment}
 \mathcal{L}_3(\mathbf{w}) =  \mathcal{L}_2(\mathbf{w}) +  \lambda_{D} (\mathbf{w}^T M_D \mathbf{w})/ |\mathbf{w}|^2
\end{equation}

In the remainder of this section, we explain how this regularization term relates to the original high-level idea of learning $\mathbf{w}$ from a larger sample and how $M_D$ is computed. We do not know the labels, $y_D$, but we do know the true distribution of the vectors $X_D$, and thanks to user feedback we have a small set $X_t, y_t$ of true labels. Label propagation \cite{Zhu2002-hd,Zhu2009-xp} is a semi-supervised learning algorithm that takes the above as inputs and generates a soft (non-binary) approximation $\hat{y}_D$ of $y_D$ given $X_D$ and $y_t$.  

The high-level assumption of the propagation algorithm is that similar points in $X_D$ should have similar values of $y_D$. Implicitly, this leads to high-density clusters in $X_D$ having homogeneous values. 
 
Operationally, label propagation requires using a $k$-nearest neighbor graph of elements in $X_D$, which we compute using an implementation of NN-descent\cite{Dong2011-hs}, an approximate but scalable way to compute a kNN graph over large datasets. We can represent this kNN graph  by its adjacency matrix $W$.  $w_{ij}$ in the adjacency matrix is a similarity score  $x_i$ and item $x_j$ in the database.  Following \cite{Zhu2002-hd} we use $w_{ij} = \exp{(-(X_i - X_j)^2/2\sigma^2)}$ as our similarity metric i.e., we let the similarity metric decay exponentially with the distance between embedding vectors decreases. $\sigma$ is a scalar hyper-parameter controlling how fast the similarity metric drops.  The propagation algorithm in~\cite{Zhu2002-hd} minimizes the total differences between neighboring vertices in the kNN graph: $\sum_{i,j} w_{ij} (y_i - y_j)^2$.  As explained in \cite{Zhu2002-hd}, this sum can be stated concisely as:
\begin{equation}
\label{eq:matmin}
 y^T (D - W) y
\end{equation}

\revision{where $W$ is the adjacency matrix, and $D$ is a closely related ``degree'' matrix, a diagonal matrix where each diagonal entry is the sum of the corresponding row in $W$, both matrices are derived from computing a kNN graph for the vectors in $X_D$.   Both $D$ and $W$ are square matrices of size $N \times N$ where $N$ is the size of the database $X_D$. In practice, $W$ is large but sparse because it only has $k$ non-zero entries per row, one for each of the $k$ neighbors}

\vspace{.1in}
\noindent {\bf DB alignment approximation}. We can obtain \autoref{eq:db_alignment} by observing the final $\mathbf{w}$ will be fitted to these synthetic labels $\hat{y}_D$ using logistic regression, which fits a sigmoid to the data. Because we empirically observe linear models fit CLIP embeddings well (\autoref{fig:locality}), i.e., concepts are clustered, then it is reasonable to assume $\text{sigmoid}(X_D \mathbf{w} ) \sim y_D$ in practice,  and therefore, we can replace $y$ in Equation~\ref{eq:matmin} with $y(\mathbf{w}) \coloneqq \text{sigmoid}(X_D \mathbf{w} )$, to obtain $y^T(\mathbf{w}) (D - W ) y(\mathbf{w})$.  Instead of one minimization to find $\hat{y}$ and a separate one to find $\mathbf{w}$, we can now add a term $y^T(\mathbf{w}) (D - W ) y(\mathbf{w})$  to \autoref{eq:clip_alignment} and perform a single minimization with respect to $\mathbf{w}$. 
\revision{The expression $D-W$ is an $N\times N$ sparse matrix, where at most $k\times N$ entries are non-zero, and where $N$ is the size of the database. $X_D$ is of size $N \times 512$, where 512 comes from CLIP, and the vector $y_D$ is of size $N$. It is easy to see that this unified approach can be slow because it scales with $N$, the size of the dataset.}

However, we can bypass this blow-up problem if we replace $\text{sigmoid}(X_D \mathbf{w} )$ with $X_D \mathbf{w}/|\mathbf{w}|$, yielding $\mathbf{w}^T X_D^T \left( D-W \right)X_D \mathbf{w}/|\mathbf{w}|^2$. 

$\mathbf{w}^T X_D^T \left( D-W \right)X_D \mathbf{w}/|\mathbf{w}|^2$ can be interpreted on its own right as penalizing drastic variation of the cosine score in highly dense regions of the graph. Normalizing $\mathbf{w}$ is meant to avoid $\mathbf{w}$ being pulled to 0 in order to make the expression 0. Because the derivative of the cosine similarity between vectors is minimized when the value of the cosine is 1, this term points $\mathbf{w}$ toward the center of a dense region instead of its periphery when either direction explains the few labeled samples equally well.

We define $M_D = X_D^T \left(D-W\right) X_D$, grouping the matrix in the middle of the expression.  $M_D$ can be computed during dataset pre-processing by building a kNN graph, which gives us both $\mathbf{w}$ and $D$,  as we do for propagation, and then computing the product $X_D^T \left(D-W\right) X_D$. This product is computed only once at preprocessing.  \revision{If this preprocessing cost becomes prohibitive, we found that using a sample of a few thousand vectors from $X_D$, instead of the full $X_D$ database, produces a very similar $M_D$ (note that we did not enable this optimization in our experiments).}

We note that it makes sense to ask: if we had access to these propagated labels $\hat{y}_D$, we could also use these propagated labels directly as a score, instead of now fitting a linear $\mathbf{w}$. However, fitting a $\mathbf{w}$ as we do above is not only a runtime optimization, it also increases accuracy. The $\hat{y}_D$ do not work as well in practice as the fitted $\mathbf{w}$. As we saw in \autoref{fig:locality}, a linear model is a good description for many queries, reflected in the low error for the best-fit linear models, which suggests restricting the model to be linear may work in our favor.

\subsection{Multiscale Representation}
\label{sec:multiscale}
In this section we describe the multi-scale representation. In the evaluation, we show this basic technique can greatly help zero-shot CLIP searches on their own,  as well as when combined with CLIP and DB alignment.  Multi-scale representation maps images to multiple vectors, increasing the size of the vector database. While this is not an issue for the vector database we use, it can be an issue for techniques that scale poorly with the size of the database. Because CLIP and DB alignment scale with the size of the data seen by the user, it is possible to combine them while keeping latency interactive.

The CLIP image embedding model is trained on images of $224 \times 224$ pixels.  However, in real-world datasets, and also those from the COCO, LVIS and BDD datasets in our evaluation, the raw images are typically larger, in the $800 \times 1000$ range.   The simplest option to use CLIP with these datasets is to rescale images to fit within this window, which we will call a {\em coarse} embedding. CLIP itself was trained this way.  This coarse embedding approach is how CLIP is most commonly used. When analyzing the types of queries where \sys/ performs poorly, however, we found this coarse approach misses many objects depending on the object size within the image.  For example, wheelchairs and animals often occupy just a few tens of pixels in dash-cam images from the BDD dataset.
\paragraph{{\bf Multi-scale patches}} An alternative to the coarse embedding approach is to treat images as a tiling of image patches at multiple size scales. Each patch is then encoded separately using CLIP, yielding multiple vectors per image. In our experiments with this multi-scale representation, we used the simplest possible combination of scales: a large-scale patch covering the full image, i.e., the coarse embedding, plus a finer-grained tiling of 1/2 the size of the image, as long as the resulting patch was larger than 224 pixels.  For example, an image of size  $448 \times 448$ maps to one coarse tile of size $448 \times 448$, plus 9 finer-grained tiles of size $224 \times 224$, corresponding to a patch of size $224 \times 224$ striding the image with a stride length of 224/2, i.e., a 10x increase in vectors per image. A smaller image would only map to one vector. A larger square image would still only include 9 vectors, though a wider image may add more along that dimension. 

At query time, an image's score is computed as the maximum score of any of its patches. This choice means \sys/ uses the vector store to find high-scoring patches, rather than high-scoring images, helping \sys/ return results where only a part of the image is relevant to the final result, and its individual benefit is shown in the evaluation.

To integrate the multi-scale representation with the user's box annotations, the region boxes corresponding to patches on the image that we have indexed ahead of time are compared to the box patches drawn as feedback by the user. Boxes that overlap with the user feedback are considered positives, and boxes with no overlap are considered negatives for the purposes of creating a training set $X_t,y_t$ for the next round.

\subsection{Solving for \texorpdfstring{$\mathbf{w}$}{\mathbf{w}}}

\begingroup

\setlength{\tabcolsep}{6pt} 
\renewcommand{\arraystretch}{2} 
 \begin{align} 
 \label{eq:total_loss} 
& \mathbf{q}_{t+1} = \argmin_\mathbf{w} \mathcal{L}(\mathbf{w}, \mathbf{x}_{i=1}^t, y_{i=1}^t) && 
\end{align}

\begin{table}[h!]
    \centering
    \begin{tabular}{lr}
  $\mathcal{L}(\mathbf{w}, \mathbf{x}_{i=1}^t, y_{i=1}^t) \coloneqq $ &   \\  
        $ \sum_{i=1}^{t}  \text{LogLoss}\left(y_i, \text{sigmoid}(\mathbf{w}^\top \mathbf{x}_i+b ) \right) $ & {\em fit user feedback} \\
         $+ \lambda 
    		  |\mathbf{w}|^2 $ & {\em but avoid $|\mathbf{w}| \to \infty$} \\
        $+ \lambda_{\text{text}} 
    		 \left( 
    		1 - \dfrac
    		{ \mathbf{w}^\top \mathbf{q}_{\text{text}} } 
    		{ |\mathbf{w}| | \mathbf{q}_{\text{text}}| }
    		\right) $ & {\em prefer $\mathbf{w}$ aligned with $\mathbf{q}_{\text{text}} $} \\
        $+ \lambda_{\text{DB}}  \left( \dfrac{\mathbf{w}}{|\mathbf{w}|}^\top \mathbf{M}_{\text{D}} \dfrac{\mathbf{w}}{|\mathbf{w}|} \right) $ 
        & {\em prefer $\mathbf{w}$ aligned with DB } 
    \end{tabular}
    \caption{Loss function}
    \label{eq:loss}
\end{table}

\endgroup


  The full loss function is written down in \autoref{eq:total_loss} and \autoref{eq:loss}

We minimize $\mathcal{L}$ using the PyTorch \cite{Paszke2019-mp} implementation of the L-BFGS \cite{Liu1989-mz} optimization algorithm to solve for $\mathbf{w}$, then we use the solution vector as the next query $q_{t+1}$ into the vector database. 

L-BFGS finds the optimal solution in a few tens of steps (taking a few milliseconds) by using second-order derivative information in addition to gradients. We use it for \sys/ because it converges quickly and also removes the need for learning rate tuning (and also the possibility of divergence or no convergence). Note in particular that the loss function computations grow with the amount of feedback the user provides, not with the size of the database.


\section{Evaluation}
\label{sec:evaluation}

The main goals of this evaluation are:
\begin{enumerate}
\item To compare \sys/ to zero-shot CLIP, which uses clip alone and no feedback, few-shot CLIP, which uses user feedback to train a logistic regression model, and a state-of-the-art active learning search baseline which also leverages user feedback: Efficient Non-myopic Search (ENS) \cite{Jiang2017-iy}. We find  \sys/ is able to show consistent improvement  in Average Precision (AP) across four datasets and thousands of search queries, in contrast to few-shot CLIP and ENS, which show consistent drops in AP with respect to zero-shot CLIP (\autoref{sec:accuracy}). We discuss how the ENS accuracy drop is partly due to a stronger reliance on ranking scores being well calibrated as probabilities in \autoref{sec:baselines}, something CLIP does not provide.
\item To show how the different \sys/ components: CLIP alignment, query alignment, and multiscale representation interact and contribute to the overall \sys/ results in (\autoref{sec:breakdown}).

\item To show \sys/ translates these algorithmic insights to time savings for users~(\autoref{sec:user}).
\end{enumerate}

\subsection{Accuracy Benchmark} 
\label{sec:accuracy}

The ideal benchmark datasets and queries for automated testing of a system like \sys/ would be varied large-scale image datasets from different domains. The images would not be from the open web, as CLIP and similar models are likely to have been trained on them.  The datasets would include labels for thousands of queries of different complexity. Results for these queries within the datasets should be both rare in an absolute sense and, more importantly, hard to find with zero-shot CLIP alone. A smaller subset of hundreds of queries where zero-shot CLIP performs very well would be included to flag regressions caused by different proposed methods.  The annotations for each query would include the relevant region within the image in order to let region-based feedback algorithms approximate human input for each image.

Because such a benchmark does not exist and creating it would be extremely costly and laborious, we opt instead to adapt several  object detection and classification  datasets:  LVIS\cite{lvis_dataset},  ObjectNet\cite{Barbu2019-gw},  BDD\cite{bdd}, and COCO\cite{coco}.  These labeled object detection datasets allow us to use their object annotations for testing region feedback, as well as for evaluating results.

\paragraph{\bf Benchmark task} For a given category and dataset, our benchmark task consists of finding 10 examples of the category,  starting with the category name string as the initial text query. We stop at 60 images if 10 examples have not been found by then. 60 images correspond to bounding searches to less than 5 minutes based on our measurements that show users may take up to 5 seconds per image (\autoref{tab:user_latency}), a quantity that will depend on the dataset and amount of box feedback provided. Though the exact cutoff choices of 10 and 60 are not critical for the results we report, having some kind of cutoff at low values ensures the reported metrics do not reflect improvements attainable only after receiving a large number amount of feedback or improvements which are only visible deep into the list of results.

\paragraph{\bf Average Precision (AP)} We measure result quality through AP.  As discussed earlier, AP for a single query is the average of the individual precision scores computed at every possible recall threshold: $AP=(\sum_{i=1}^R P_i)/R$, where $R$ is the number of relevant results in the data and $P_i$ is the precision if we cut results off after the $i^{th}$ relevant result in the data. Because we consider up to 10 relevant results, only ten precision scores will be included in the average.  For queries with R < 10 we use R instead.  When less than the target number of results are found, the remaining precision scores are set to 0.  AP ranges 0 to 1, with 0 meaning no results were found within 60 images, and 1 meaning the first ten images were all positive. In addition to stopping the benchmark at 60 images, within those 60 images AP favors earlier results rather than later ones.

\paragraph{\bf Zero-shot CLIP results} The benchmark task simulates a scenario where a user starts a search query via text and then interacts with it via region box feedback. For a given dataset category such as wheelchair, we use the text string ``a wheelchair'' to start the search. After receiving the first image result, the benchmark code uses the dataset ground truth to determine when the image is relevant, and then provides box labels from the dataset as region based feedback around the relevant image area. For zero-shot CLIP this feedback is ignored, and we obtain the results shown in \autoref{fig:zero_shot_variance}.

Note the long left-tails in the figure. ObjectNet and LVIS, in particular, help broaden this evaluation thanks to their large number of categories (respectively 300 and 1400), and a corresponding large number of queries on the left tails. COCO and LVIS, like many other standard benchmarks in object detection, share the same underlying set of images taken from the open web (Flickr), which suggests CLIP is likely to have seen the images during training. LVIS includes annotations for many objects besides the main subject of their images, in contrast to COCO, which is reflected in COCO having almost no density at lower accuracies. 
BDD consists of a small number of classes from driving scenes, and shows a similar problem of low density because most of the classes (car, person, etc.) are ones CLIP was likely trained extensively on.

These observations suggest even datasets with high mean AP on standard labels in practice have hidden long-tails when it comes to ad-hoc queries.
Because average metrics over queries in a dataset are dominated by the high-density regions where CLIP performs well, but we are interested in ``long tail" queries in \sys/,  we use a cutoff line at $AP < .5$ to define a  hard subset of the queries, marked in the plot by the dashed line.  The labels on the dashed line indicate the fraction and total number of classes in the dataset that are below this threshold.

COCO and BDD are designed to test object detection search, hence many labeled classes are not rare in absolute terms, which makes their left tails much less informative.  However they are larger datasets both in image size and number of images, so they help test the latency of different algorithms. 

ObjectNet only includes images of size of size $224 \times 224$, the same size used to train CLIP, with intentionally centered objects. This dataset feature could push baseline CLIP to perform better than it would otherwise, but ObjectNet still challenges CLIP for a several tens of queries.
\subsection{\sys/ Benefits}
In later sections we show zero-shot CLIP performs better on average than the other baselines we evaluate. Hence, in this section we show that \sys/ beats zero-shot CLIP, especially on classes that CLIP struggles with, by  running \sys/ on the benchmark task defined previously and quantifying the distribution of the change in AP ($\Delta \text{AP}$), rather than only the average change.

The hyperparameters of \sys/ for this experiment are $k=10$ for the kNN graph, $\sigma = .05$ for the distance kernel used between vertices in the graph, $\lambda = 100$ for the norm regularization, $\lambda_c = 10$ for clip-alignment regularization, and $\lambda_D = 1000$ for database-alignment regularization.  We chose these parameters based on average performance on the queries. We note the same hyperparameters were used for all queries and datasets, and the different datasets seem to peak at the same parameter values. Varying these $\lambda$ one order of magnitude decreased the absolute accuracy slightly, but \sys/ remains substantially more effective than the baseline. Varying $k$ from 5 to 20 also did not substantially  affect results.

\begin{figure}[ht]
\caption{CDF of change in AP of \sys/ on top of zero-shot CLIP, broken down by dataset.  The red shade marks the region with $\Delta \text{AP} < 0$.  The darker gray shaded area marks the $[.1,.9]$ quantile interval of observed $\Delta AP$ so that the robustness of the improvement is evident. The solid vertical lines mark the minimum, median and maximum change observed for completeness. The average values are shown in \autoref{tab:breakdown} }
\label{fig:quantiles}
\includegraphics[width=3.3in]{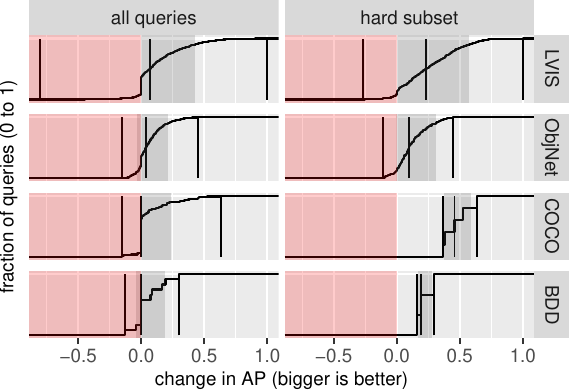}
\end{figure}

\autoref{fig:quantiles} shows $\Delta \text{AP}$ broken down by dataset (left column).
Note that zero-shot CLIP reaches AP of 1 (see \autoref{fig:zero_shot_variance}) for several object classes, which guarantees $\Delta \text{AP} \leq 0$  for those queries, reflected by the vertical jump at $\Delta \text{AP}=0$ in the CDF.
In order to highlight the change on the tail queries, we show the distribution of $\Delta \text{AP}$ restricted to the hard subset of queries on the right column. A lower starting AP allows $\Delta \text{AP}$ to be generally larger.

Here, the gray shaded area marks the $[.1,.9]$ quantile interval. The red-shaded region (AP<0) shows queries where \sys/ did not improve performance. In general, \sys/ is quite robust, with more than 90\% of the queries improving or staying the same (we a breakdown of average AP results on these datasets in the next section).
The solid bars plot the min, median and max $\Delta \text{AP}$. In many cases the min is very close to 0. The main reason for a few outliers on the left seems to be the multiscale representation, which in most cases is beneficial, but can sometimes push the first result of a query a few spots down the rank. This can affect AP strongly if the original AP was 1. 
Methods to reap the benefits of multiscale while limiting any downside are an interesting line of future work.

\subsection{\sys/ Breakdown}
\label{sec:breakdown}



In this section we breakdown the contributions from the different \sys/ components to the overall accuracy in the benchmark presented before.

\autoref{tab:breakdown} shows how much each of these optimizations contributes to the total Average Precision (AP) in \sys/ by adding optimizations one at a time and recording the aggregate changes in AP. It shows increases in mean AP for each optimization (rows), on each dataset (columns). The row states the mean AP when using CLIP embeddings out of the box. Each later row adds one of the optimizations in \sys/.
The rightmost column ``avg'' is the average of the values to the left. It shows all optimizations contribute to the final results on average. The rest of the columns show the mean AP for each dataset, and each shows different benefits from different optimizations in \sys/.  Each row represents the addition of a different optimization described in the indicated section of the paper.   Zero-shot CLIP is CLIP without any user input; few-shot CLIP is CLIP (with multiscale, in this case)  plus logistic loss from user feedback (Equation~\ref{eq:lr}), which is also a baseline approach to relevance feedback on its own right.

The exact mean Average Precision (mAP) values mostly reflect the high initial mAP of zero-shot CLIP for many queries, as explained in \autoref{fig:zero_shot_variance}, here focus more on the relative sizes of the changes, which may transfer better to ad-hoc scenarios. Every row of optimization consistently increases mAP with the exception of few-shot CLIP, which causes a drop on most columns. We referenced this phenomenon in the introduction, as few-shot CLIP is a reasonable,  baseline for relevance feedback. We note few-shot CLIP when combined with alignment methods undo this regression. The bottom row is equivalent to total \sys/ mAP. 

Few-shot CLIP does improve the mean AP for the hard subset of queries in LVIS, something we observe consistently on the hard subset of queries when multiscale is off (not shown in this table).  Because zero-shot CLIP is very accurate, the weakness of the few-shot approach is more apparent on the top half of the plot and motivates CLIP alignment, which  turns out to also offer strong benefits in hard queries where few-shot CLIP improves results, as in LVIS.  When the zero-shot CLIP vector performs the worst at finding initial results we may be the most tempted to ignore that vector weight and prefer the data, but this would be a mistake in the setting of small samples because the generalization error of the learned vector is likely dominated by {\em variance} error, whereas the zero-shot CLIP vector error consists of {\em bias} error.

Finally, we highlight the importance of algorithms that can scale to handle the growth in vectors that multiscale brings, as this brings has strong benefits for the search. In \autoref{tab:breakdown} these rank second only to CLIP align, especially on BDD which has the largest images, but also on LVIS, which includes many objects on different image locations. ObjectNet does not benefit from multiscale, as it consists of standard fixed size images. Additionally, multiscale interacts with CLIP align in more complex ways than just latency: CLIP align and DB align are less effective in absolute terms when multiscale is not enabled. The data on \autoref{tab:baselines}, which we will explain fully later, includes a row for \sys/ without multiscale enabled. Especially on BDD, the 3 hard queries improve from .02 to .07 without multiscale, but from .10 to .24 with it.

\begin{table}[ht]
\caption{Increases in mean Average Precision (mAP) for each optimization (rows), on each dataset (columns). The top panel shows the averages over all queries, while the bottom panel is restricted to the hard subset of queries for each dataset. All techniques contribute to the final result, their relative contribution depends on properties of the data.}
\label{tab:breakdown}
\begin{tabular}{l|lrrrr|r}
\toprule
 &  & LVIS & ObjNet & COCO & BDD & avg. \\
\midrule
\multirow[c]{5}{*}{\parbox{.3cm}{\rotatebox[origin=c]{90}{all queries}}} & zero-shot CLIP & 0.63 & 0.64 & 0.90 & 0.74 & 0.72 \\
\cline{2-2}
& +multiscale (\autoref{sec:multiscale}) & 0.70 & 0.64 & 0.95 & 0.76 & 0.76 \\
\cline{2-2}
& +few-shot CLIP & 0.67 & 0.59 & 0.87 & 0.68 & 0.70 \\
\cline{2-2}
 & +Query align (\autoref{sec:clipalign}) &  0.75 & 0.69 & 0.96 & 0.77 & 0.79 \\
\cline{2-2}
 & +DB align (\autoref{sec:dbalign}) & 0.76 & 0.70 & 0.96 & 0.79 & 0.80 \\
\cline{1-2} \cline{2-7}
\multirow[c]{5}{*}{\parbox{.3cm}{\rotatebox[origin=c]{90}{hard subset}}} & zero-shot CLIP & 0.19 & 0.28 & 0.27 & 0.02 & 0.19 \\
\cline{2-2}
& +multiscale & 0.32 & 0.28 & 0.58 & 0.10 & 0.32 \\
\cline{2-2}
& +few-shot CLIP & 0.34 & 0.28 & 0.57 & 0.07 & 0.31 \\
\cline{2-2}
 & +Query align & 0.42 & 0.39 & 0.74 & 0.20 & 0.44 \\
\cline{2-2}
 & +DB align & 0.44 & 0.40 & 0.75 & 0.24 & 0.46 \\
\bottomrule
\end{tabular}
\end{table}

\subsection{Comparison with  Baselines}
\label{sec:baselines}
\begin{revisionenv}
    We compare \sys/ to the following baselines: ENS from the recent active learning literature and Rocchio's algorithm,\cite{rocchio71}, a classic relevance feedback algorithm. 
\end{revisionenv}

\begin{revisionenv}
    
\paragraph{\bf Rocchio's algorithm\cite{rocchio71, irbook}} Rocchio's algorithm creates a new query vector at every iteration, $q_t$, by taking the initial query vector used, $q_0$ and adding the average of the set of relevant example vectors returned so far ($D_r$) (with a weight of $\beta$, and subtracting the average of the nonrelevant example vectors seen so far($D_n$) (with a weight of $\gamma$). The formula we use is \autoref{eq:rocchio}:

\begin{equation}
\label{eq:rocchio}
\mathbf{q}_n = \alpha \mathbf{q}_0 + \frac{\beta}{|D_r|} \sum_{d_j \in D_r} \mathbf{d_j} 
- \frac{\gamma}{|D_n|} \sum_{d_j \in D_n} \mathbf{d_j}
\end{equation}

Like \sys/, Rocchio's algorithm includes hyperparameters that weight the three terms. For the experiments below, we use the hyperparameters that maximized the average AP across all datasets: $\beta = .5$ and $\gamma = .25$. Additionally, following \cite{irbook}, we tested $\gamma = 0$, but AP was higher with our choice of $\gamma = .25$. $\alpha$ was set to 1 as any other value would be equivalent after rescaling.

\end{revisionenv}

\paragraph{\bf Efficient Non-myopic Active Search (ENS) \cite{Jiang2017-iy}} ENS is a state-of-the-art baseline from the active learning literature. Unlike much active learning work that aims to optimize the accuracy of a model by judiciously choosing data points to label,  ENS is an {\em active search} algorithm whose goal is to maximize the number of results found for a fixed amount of human input, which is similar to the setting of \sys/.  ENS has been shown to be more efficient than previous work in the area such as \cite{Garnett2012-tp}.  

 While both \sys/ and zero-shot CLIP  select the next result greedily on every iteration, ENS takes a long view: it picks the next result based on its estimate of how this choice affects the total number of results found after $t$ steps in the future (in the benchmark case, $t=60$). For example, ENS may pick the second-highest scoring image now if it lies within a dense cluster of similarly scored points. In that scenario, if the image turns out to be relevant then ENS can estimate it is more likely to find many more results in succession. This choice could be wiser than picking the highest scoring vertex in the kNN graph if this result is isolated. Zero-shot CLIP would not consider this scenario. \sys/ would pick the highest scoring vector just like zero-shot CLIP, though DB alignment will affect the scores of isolated points.
 
 We implemented the ENS algorithm in \sys/.  ENS makes use of a kNN graph, the same one we used for DB alignment. For its hyper-parameters we used $k=20$ since that improved ENS results, and we used the same $\sigma=.05$ as for \sys/. We changed our version of ENS to integrate CLIP into it: we use CLIP scores as an individual $\gamma_i$ for each vertex so ENS also has access to CLIP as a prior. The $\gamma$ parameter acts as a prior score in ENS, providing a score for vertices without labeled neighbors. The original ENS uses a global $\gamma=.1$ hyperparameter. Additionally, we wait for zero-shot CLIP to find a first positive result from the data before starting ENS. Both of our modifications help ENS perform better in the benchmark. For simplicity,  for this benchmark we set the time horizon $t= 60$ initially, and reduce it after every step so ENS can make optimal decisions given the time remaining.

\paragraph{\bf Baseline results} We ran the benchmarks described in \autoref{sec:accuracy},  including zero-shot CLIP, \sys/,   and show the mAP results in \autoref{tab:baselines}.  In the table, each row corresponds to one of these methods and each column to a different dataset. The rightmost column is the average of those to the left and shows that \sys/ increases mAP further.  As we do in \autoref{tab:breakdown}, we show averages over both the hard subset and all queries.  The hard subset numbers tend to spread out wider, which is more helpful to display differences, \revision{We note \sys/ AP is highest for all datasets in this long tail of hard queries. Rocchio comes in second, and is slightly better than \sys/ for COCO when aggregating across all queries. This behavior is consistent with Rocchio also using a form of CLIP alignment through the $q_0$ term. We note that few-shot CLIP generally lags behind Rocchio in AP, and for some datasets is also worse than zero-shot CLIP, showing the importance of regularizing by the initial query, done implicitly in Rocchio and explicitly in \sys/.}  \autoref{tab:baselines} also shows ENS can decreases AP with respect to zero-shot CLIP.  
Note that because we only implemented ENS for coarse embedding, without multiscale, in \autoref{tab:breakdown} we compare all baselines without multiscale enabled. \sys/ with multiscale enabled was evaluated in \autoref{sec:breakdown}.


\begin{table}
\caption{Table comparing mean Average Precision (mAP) of \sys/ different baseline algorithms. No method used multiscale in this table. }
\label{tab:baselines}
\begin{tabular}{l|l|rrrr|r}
\toprule
 &  & LVIS & ObjNet & COCO & BDD & Avg. \\
\midrule
\multirow[c]{3}{*}{\parbox{.9cm}{\rotatebox[origin=c]{90}{\parbox{.9cm}{\centering all queries}}}}  
& zero-shot CLIP  & 0.63 & 0.64 & 0.90 & 0.74 & 0.72 \\
\cline{2-2}
& \revision{few-shot CLIP} & \revision{0.65} & \revision{0.58} & \revision{0.88} & \revision{0.73} & \revision{0.71} \\
\cline{2-2}
 & ENS\cite{Jiang2017-iy} & 0.50 & 0.43 & 0.86 & 0.70 & 0.62 \\
\cline{2-2}
 & \revision{Rocchio\cite{rocchio71}} & \revision{0.68} & \revision{0.70} & \bfseries \revision{0.93} & \revision{0.75} & \revision{0.76} \\
\cline{2-2}
 & this work & \bfseries 0.69 & \bfseries 0.70 & 0.92 & \bfseries 0.76 & \bfseries 0.77 \\
 
\hline
\multirow[c]{3}{*}{\parbox{.9cm}{\rotatebox[origin=c]{90}{\parbox{.9cm}{\centering hard subset}}}} & zero-shot CLIP  & 0.19 & 0.28 & 0.27 & 0.02 & 0.19  \\
\cline{2-2}
& \revision{ few-shot CLIP} & \revision{0.25} & \revision{0.28} & \revision{0.32} & \revision{0.06} & \revision{0.23}
\\ 

\cline{2-2}
 & ENS & 0.16 & 0.24 & 0.37 & 0.03 & 0.20 \\
\cline{2-2}
  & \revision{ Rocchio} & \revision{0.28} & \revision{0.38} & \revision{0.49} & \revision{0.05} & \revision{0.30} \\
\cline{2-2}
 & this work & \bfseries 0.30 & \bfseries 0.40 & \bfseries 0.55 & \bfseries 0.07 & \bfseries 0.33 \\
\bottomrule
\end{tabular}
\end{table} 

\paragraph{\bf ENS analysis} One reason for the drop in ENS is that the ENS logic to estimate the expected reward is sensitive to score calibration. Calibrated scores need to be correct not only in the ranking they produce but also correct when interpreted as probabilities: for example, 10\% of points with a calibrated score of .1 should be positive.  Calibration is a strong assumption; we note that CLIP scores are not calibrated.  


ENS depends on score calibration because it uses the probabilities as weights to compute expected values. The longer the reward-horizon hyper-parameter $t$, the more terms this sum has and the more susceptible the score is to calibration errors. We test this hypothesis by calibrating CLIP scores $s_i$ for each vector into probabilities via Platt scaling \cite{Niculescu-Mizil2005-hg,Platt2000-tc}. We emphasize this calibration is not possible in a real deployment because it requires labeled data ahead of time. Then we run ENS using this carefully tuned $\gamma_i$, and show the average of mAP over all the datasets for of ENS without $\gamma_i$ calibration on the top row of \autoref{tab:horizon}, and those with calibrated $\gamma_i$ in  the second row.

The table shows ENS performs better if the initial $\gamma_i$ are calibrated using Platt scaling, with access to the ground truth data, showing ENS is sensitive to calibration.  Furthermore, mAP degrades sharply as we increase the reward horizon parameter used internally by ENS to compute expected values, whereas it degrades less sharply in the bottom row, showing that larger reward horizons are more sensitive to poor calibration. In contrast, greedy methods such as zero-shot CLIP do not depend on the calibration of the scores. \sys/, in particular, optimizes dot product alignment in \autoref{eq:db_alignment} rather than probability estimates.

\begin{table}
\caption{Mean AP scores for ENS averaged over the four datasets, showing the effect of varying the time horizon hyperparameter (columns), and the effect of calibrating the initial $\gamma_i$ scores into probabilities (second row). The results of the calibrated row are not attainable in practice because calibration requires labeled data.}
\label{tab:horizon}
\begin{tabular}{lrrrr}
\toprule
\parbox{3.4cm}{\raggedleft reward horizon $t=$} & 1 & 2 & 10 & 60 \\
\midrule
raw $\gamma_i$  & 0.63 & 0.62 & 0.61 & 0.55 \\
calibrated $\gamma_i$   & 0.65 & 0.65 & 0.65 & 0.63 \\
\bottomrule
\end{tabular}
\end{table}

A second reason for the mAP drop in ENS is that the kNN classifiers used internally for ENS are less efficient at learning from small samples than the linear models used by \sys/, as long as a linear model is a good reflection of the data, which is true for CLIP embeddings. This is reflected in \autoref{tab:horizon} on the column for time horizon $t=1$, an extreme setting where ENS effectively becomes a greedy kNN-model.

\subsection{End-to-End Tests}
\label{sec:user}

The primary goal of this section is to test \sys/ with real people in different scenarios to how different variables affect the overall time it takes to complete a task with \sys/ or otherwise. The overall time includes not just the time savings due to increased accuracy but also the time costs due to user feedback latency as well as any possible system latencies. A secondary goal is to show estimates of how much time it takes a user to provide feedback to \sys/, and how this time varies with the type of annotation. 

The effect of annotation time overhead from using \sys/ is ameliorated by two factors: the first is that users only annotate positive examples, and in hard searches that require inspecting many images, positive examples are rare. Conversely, when positive examples are common, \sys/ overhead due to labeling is larger, but the searches themselves complete much more quickly.

The baseline for this evaluation is the same user interface and back-end of \sys/, but with all the optimizations of \autoref{sec:approach} disabled, which means zero-shot CLIP with a User Interface (UI) for searching.

In this baseline system,  \sys/ provides users with a keyboard binding to mark the whole image as relevant. This input is necessary to ensure that users know when they completed the task (the system shows a total count), but is not used as feedback.
We recruited 20 graduate students and 20 Amazon MTurk workers for these tasks. Both groups were assigned the same mix of tasks on both systems.
 
 Like our task in \autoref{sec:accuracy}, we ask users to find 10 examples of the given concepts and we stop each task after 6 minutes. Note the evaluation in \autoref{sec:accuracy} involved no real users -- instead we use dataset ground truth to provide feedback and we stop the query after finding 10 examples or going through 60 images. Instead of measuring AP, in this evaluation we measure the {\em time} it takes users to complete the task (find 10 examples).

Because human time is a very limited resource, in this section we evaluate \sys/ and the baseline on 7 queries. Unlike in \autoref{sec:accuracy}, our goal in this experiment is not to be comprehensive, but to observe how searches play out in different scenarios.  Guidance on how to use, including how to provide feedback and the keyboard bindings of the UI was provided through a tutorial screen followed by two initial queries used for training users on each version of the system, and whose results were discarded.  Each user then completed seven tasks using both interfaces.  The subset of queries each user completed on each interface did not overlap, as in early tests we found the time it takes to provide feedback or skip an image can be artificially low the second time the same image is seen for the same query.

\begin{figure}[ht]
\caption{Comparing time elapsed before users find 10 examples of each of the categories, or 6 minutes pass (rightmost on $x$). A value of 360 seconds corresponds to not complete the task. The median time for each method is represented by a triangle.  The error bars mark the $95^{th}$ percentile CI for the mean time, to show variation between users. }
\label{fig:querytime}
\includegraphics[width=3.3in]{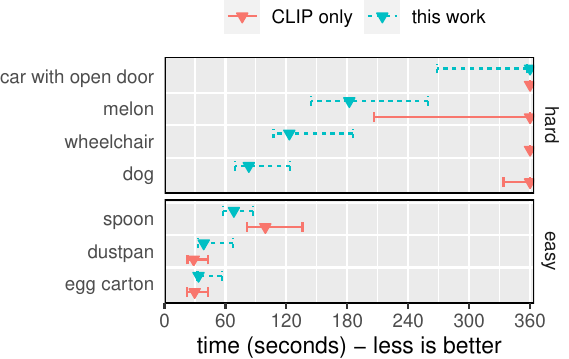}
\end{figure}

We show aggregate results for these experiments in \autoref{fig:querytime}, where the $x$ axis measures the time elapsed and ends at 6 minutes (360 seconds).  The 7 queries we tested are logically grouped into two sets: one set of ``hard'' queries and one of ``easy'' queries, which in this context simply means queries where the zero-shot CLIP has a low or high AP accuracy. The main goal of this section is to show how this affects time. 

We show the results for the baseline and \sys/ side by side distinguished by color and dashes.  The middle triangle corresponds to the median time taken by users.  The error bars on the plot correspond to a bootstrapped 95\%-CI of the time required, capturing inter-user variability for a given query.

For the hard queries, the median time for the baseline is 360 seconds, ie. half or more of the users could not complete the task within that time limit, a qualitative difference with \sys/. 

For ``wheelchair'' and ``car with open door'' none of the baseline users were able to complete the task, so the corresponding error bars collapsed into the 360 mark.
The most challenging query, ``car with open door,'' was completed only by a few people, all using \sys/, though most people, even most of those using \sys/, were not able to complete it.
The times required to find wheelchairs and dogs show \sys/ helped substantially shorten times in these cases, even after accounting for inter-user variability.

For easy queries, the figure shows that \sys/ can be slower than the baseline; this difference boils down to the annotation overhead per frame.  \autoref{tab:user_latency} shows  when images are marked as relevant the  overhead for selecting the region bounding box adds about a 50\% latency (4.5 vs. 3 seconds per image). For these queries, \sys/ adds the worst possible relative overhead but the absolute latency values are low for both methods.
\begin{table}
\caption{User annotation time (s) per image depends on whether the image is marked relevant. (2s vs 3s). Localized feedback adds an overhead of about 1.5 seconds per image (4.5s) The $\pm$ denotes the 95\% CI. }
\label{tab:user_latency}
\begin{tabular}{lrr}
\toprule
{} &  baseline &  seesaw  \\
\midrule
not marked &      1.98 $\pm$ .10 &    2.40 $\pm$ .19 \\
marked relevant      &      3.00 $\pm$ .28 &    4.40 $\pm$ .45 \\
\bottomrule
\end{tabular}
\end{table}

\begin{revisionenv}
\subsection{System Latency and Scalability}
\label{sec:scalability}
\sys/ aims to scale to larger image datasets. One challenge when scaling to larger datasets is to keep system latency low, and system latency depends partially on the algorithms used for selecting new results based on feedback.  \autoref{tab:latency} shows system latencies as we increase the scale of the datasets. Each row corresponds to a different dataset, ordered on increasing scale (number of vectors in database). We include both the coarse and multiscale representations as different rows, as they imply differences in number of vectors. 

We highlight how \sys/ latency remains within the hundreds of milliseconds, while a propagation-based version of \sys/ shows latency increasing beyond a few seconds, one motivation for our DB alignment approximation in \autoref{sec:dbalign}.

\begin{table}[ht!]
\caption{System latency per iteration (seconds) vs dataset size (\# vectors). The extra $^-$ sign next to a dataset name means coarse indexing (one vector per image), as opposed to multiscale indexing. COCO and LVIS share the same database, so we only show one.}
\label{tab:latency}
\begin{tabular}{l|r|rrrrr}
\toprule
 & vectors & CLIP & ENS & Rocchio & \sys/ & prop. \\
\midrule
ObjNet$^-$ & 50K & 0.11 & 0.10 & 0.14 & 0.27 & 0.83 \\
BDD$^-$ & 80K & 0.09 & 0.11 & 0.10 & 0.23 & 0.90 \\
COCO$^-$ & 120K & 0.10 & 0.22 & 0.16 & 0.34 & 1.11 \\
BDD & 1.6M & 0.13 & NA & 0.16 & 0.34 & 2.95 \\
COCO & 1.6M & 0.14 & NA & 0.23 & 0.47 & 2.88 \\
\bottomrule
\end{tabular}
\end{table}

\subsection{Hyperparameters}
\label{sec:hyperparams}

\sys/ requires hyperparameters $\lambda$, $\lambda_c$, $\lambda_D$.  In this section we show \sys/ handles hyperparameter values varying an order of magnitude while still improving results vs. zero shot CLIP. \autoref{tab:hyperparam_multi_reg} shows values for each dataset, highlighting the maximum achieved. The enclosed row highlights the values we used for the evaluation benchmark. We note two things: the AP values are emparically optimal or near-optimal at similar hyperparameter values, even though the datasets are different.  

For a new dataset, we recommend starting with hyperparameter values in the same range.

{
\begin{table}[ht!]
\caption{\sys/ AP for different hyperparameter settings}
\label{tab:hyperparam_multi_reg}
\begin{tabular}{rrr|rrrr|r}
\toprule
$\lambda_c$ & $\lambda_D$ & $\lambda$ & BDD & COCO & LVIS & ObjNet & Avg. \\
\midrule
3 & 300 & 100 & 0.78 & 0.96 & 0.76 & 0.68 & 0.80 \\
3 & 1000 & 100 & 0.77 & 0.97 & \bfseries 0.77 & 0.68 & 0.80 \\
3 & 3000 & 100 & 0.77 & 0.96 & 0.76 & 0.63 & 0.78 \\
10 & 300 & 100 & 0.78 & 0.96 & 0.75 & 0.69 & 0.80 \\
10 & 1000 & 30 & 0.79 & 0.96 & 0.76 & 0.70 & 0.80 \\
\hline
\rowcolor{lightgray}
10 & 1000 & 100 & \bfseries 0.79 & 0.96 & 0.76 & \bfseries 0.70 & \bfseries 0.80 \\
\hline
10 & 1000 & 300 & 0.79 & 0.96 & 0.76 & 0.70 & 0.80 \\
10 & 3000 & 100 & 0.79 & \bfseries 0.97 & 0.77 & 0.69 & 0.80 \\
30 & 300 & 100 & 0.77 & 0.96 & 0.73 & 0.68 & 0.79 \\
30 & 1000 & 100 & 0.77 & 0.96 & 0.74 & 0.69 & 0.79 \\
30 & 3000 & 100 & 0.77 & 0.96 & 0.74 & 0.69 & 0.79 \\
\bottomrule
\end{tabular}
\end{table}
}

\end{revisionenv}

\section{Related work}
\label{sec:related}
\sys/ relates to work in the following areas:

{\bf Relevance feedback} \cite{Salton1990-ks} with explicit region annotation feedback, a common interface for images\cite{Zhou2003-hb}. \sys/'s novelty lies in the specific algorithms it uses, which make it work better than existing approaches on CLIP embeddings.  Rocchio's algorithm\revision{\cite{rocchio71}} is one of the baselines we used in this paper.


{\bf Systems for image search}, such as \cite{Vasconcelos_undated-ja}, PicHunter \cite{ Cox2000-ai}, MindReader \cite{Ishikawa1998-dx} and Falcon \cite{Wu_undated-xy}. Much of this work aims to help users query by example. CLIP enables us to start the query by using text, and we find it is important to use the CLIP query vector as a regularizer term in \ref{sec:clipalign}.
Some existing work, such as the baseline approach in \cite{Grossman2016-total_recall_trec}, use the query as a positive example.  
Unlike Falcon, and other work such as \cite{renders_2017} which aim to navigate non-convex results. However, because CLIP embeddings seem to well clustered \sys/ can focus on this specific case.

{\bf Text-image retrieval} aims to search images with text. Recent work includes Drilldown \cite{drilldown}. The goal is this work is to provide a richer interface with text-based refinement, though we are able to assume some amount of labeling on the user's data. \sys/ aims to help the tail of queries on a user's database where CLIP does not work well by leveraging user input, without requiring user labels. 

{\bf Active learning} \cite{Settles_undated-nl} also studies human-in-the-loop labeling but does not directly address the search problem of relevance feedback. {\bf Active search} \cite{garnett_active_search} is a subset of active learning techniques for searching datasets more directly related to \sys/. ENS is a state of the art of active search approach  we compared \sys/ to in our evaluation.  Active learning can often be expensive at runtime on large datasets due to linear or super-linear scaling. Recent work such as \cite{Coleman2022-ag} points out this scaling problem and instead aims to scale these approaches by restricting active learning algorithms only to the neighborhood of search queries and leveraging vector stores to explore that neighborhood. \sys/'s approach to database alignment\autoref{sec:dbalign} shows there can be some value to considering the global database structure in addition to the immediate neighborhood of the query.

{\bf Semi-supervised learning} \cite{Zhu2009-xp} proposes approaches for leveraging both labeled and unlabeled data. \autoref{sec:dbalign} in particular, can be seen as semi-supervised learning, or domain adaptation. Label propagation \cite{Zhu2002-hd} is  a semi-supervised learning technique and has been used together with active learning for search in \cite{Zhu2003-dh}. Label propagation can be expensive to scale.  Manifold regularization \cite{Belkin2006-tx}, another topic in semi-supervised learning, is closely related to both label propagation and to the regularization term we use in \autoref{sec:dbalign}. Manifold regularization by itself is expensive, just like label propagation, but collapsing the term into a small quadratic term as done in \autoref{sec:dbalign} makes it possible.

{\bf High-recall or total-recall retrieval} \cite{Grossman2016-total_recall_trec},\cite{abulsaud_high_recall_retrieval},\cite{Yu_menzies_2018},  aims to find all relevant results within a database. This is an important task in legal contexts for example, where legal discovery needs to find examples and also provide some assurance that a substantial part of the results was not missed. Some of the approaches proposed in the area are similar to the few-shot CLIP approach.  Total-recall retrieval is a harder problem than finding some positive results within the database, and may require a large amount of exploration of likely irrelevant results to provide some assurance about all regions of the database.

\begin{revisionenv}
    
{\bf Multiscale representations} The usefulness of multiscale representations of images of one form or another has been recognized for a long time \cite{Burt1983-er,Adelson1984-ch, Simoncelli1995-bm}, and internal representations used by CNN-based object detectors \cite{Lin2016-vr, Dollar2014-nt} incorporate it in different ways.  In this work, this general idea is used to overcome some of CLIPs limitations, allowing us to handle more complex images with relevant object in small portions of images. 

{\bf Query relaxation techniques}\cite{Elbassuoni2011-cw} rewrite or modify user queries by removing tokens or constraints from the query. These techniques help increase recall when initial queries are unnecessarily restrictive.

{\bf Active learning for data cleaning} \cite{Krishnan2016-aa}, aims to help users build better models by reducing the amount of data that must be cleaned. Additionally active learning has been applied to other database problems like {\bf entity resolution} \cite{Kasai2019-pu} and {\bf schema matching} \cite{Yan2013-eh}.  Here we also employ active learning, focusing on the problem of dataset labeling.

Finally, {\bf {\em Why} and {\em Why-Not}  provenance} techniques help users understand how their data relates to their query results and, conversely, why some of their data was excluded from the results (see \cite{CheneyCT09} for a general survey of many papers).  Systems like \sys/ that evolve their result set of over time using models are generally related, although the specifics of our techniques differ as we focus on CLIP query vector refinement rather than a more relational setting.  

\end{revisionenv}

\begin{revisionenv}
\section{Conclusion}

We  presented \sys/, a system to help users find objects in their image datasets that incorporates their feedback in the form of region-based annotations in order to improve their results.  \sys/ helps users leverage pretrained embeddings which have not been fine-tuned for their specific datasets due to a lack of captioned data. \sys/'s approach consists of framing this problem as a semi-supervised learning problem, combining two kinds of regularization objectives into a loss function, in order to handle the lack of labeled data.
At the same time \sys/ leverages a multi-scale multi-vector representation of images in order to handle embedding model limitations, and integrates it with the relevance feedback mechanisms. In order for this integration to be successful, \sys/ focuses on keeping latency low despite the larger dataset sizes implied by this representation. 
We benchmark \sys/ extensively on thousands of queries, showing consistent empirical advantages over an active learning baseline, a relevance feedback baseline from the information retrieval community, and practical baselines such as few-shot learning.

Interesting directions for future work include reducing or removing explicit user feedback, for example by leveraging newer generations of models such as GPT-4, as well as further taking advantage of the abilities of vision-language models such as CLIP to provide textual feedback.

\end{revisionenv}

\clearpage

\bibliographystyle{ACM-Reference-Format}
\bibliography{refs,paperpile,ir}


\begin{thebibliography}{56}


\ifx \showCODEN    \undefined \def \showCODEN     #1{\unskip}     \fi
\ifx \showDOI      \undefined \def \showDOI       #1{#1}\fi
\ifx \showISBNx    \undefined \def \showISBNx     #1{\unskip}     \fi
\ifx \showISBNxiii \undefined \def \showISBNxiii  #1{\unskip}     \fi
\ifx \showISSN     \undefined \def \showISSN      #1{\unskip}     \fi
\ifx \showLCCN     \undefined \def \showLCCN      #1{\unskip}     \fi
\ifx \shownote     \undefined \def \shownote      #1{#1}          \fi
\ifx \showarticletitle \undefined \def \showarticletitle #1{#1}   \fi
\ifx \showURL      \undefined \def \showURL       {\relax}        \fi
\providecommand\bibfield[2]{#2}
\providecommand\bibinfo[2]{#2}
\providecommand\natexlab[1]{#1}
\providecommand\showeprint[2][]{arXiv:#2}

\bibitem[\protect\citeauthoryear{Abualsaud, Ghelani, Zhang, Smucker, Cormack,
  and Grossman}{Abualsaud et~al\mbox{.}}{2018}]%
        {abulsaud_high_recall_retrieval}
\bibfield{author}{\bibinfo{person}{Mustafa Abualsaud}, \bibinfo{person}{Nimesh
  Ghelani}, \bibinfo{person}{Haotian Zhang}, \bibinfo{person}{Mark~D Smucker},
  \bibinfo{person}{Gordon~V Cormack}, {and} \bibinfo{person}{Maura~R
  Grossman}.} \bibinfo{year}{2018}\natexlab{}.
\newblock \showarticletitle{A System for Efficient {High-Recall} Retrieval}. In
  \bibinfo{booktitle}{\emph{The 41st International {ACM} {SIGIR} Conference on
  Research \& Development in Information Retrieval}} (Ann Arbor, MI, USA)
  \emph{(\bibinfo{series}{SIGIR '18})}. \bibinfo{publisher}{Association for
  Computing Machinery}, \bibinfo{address}{New York, NY, USA},
  \bibinfo{pages}{1317--1320}.
\newblock


\bibitem[\protect\citeauthoryear{Adelson, Burt, Anderson, Ogden, and
  Bergen}{Adelson et~al\mbox{.}}{1984}]%
        {Adelson1984-ch}
\bibfield{author}{\bibinfo{person}{E Adelson}, \bibinfo{person}{P Burt},
  \bibinfo{person}{C Anderson}, \bibinfo{person}{J~M Ogden}, {and}
  \bibinfo{person}{J Bergen}.} \bibinfo{year}{1984}\natexlab{}.
\newblock \showarticletitle{{PYRAMID} {METHODS} {IN} {IMAGE} {PROCESSING}}.
\newblock \bibinfo{journal}{\emph{undefined}} (\bibinfo{year}{1984}).
\newblock
\urldef\tempurl%
\url{https://www.semanticscholar.org/paper/e49793511ba203e26b99e7e81fd15a7d505b5cea}
\showURL{%
\tempurl}


\bibitem[\protect\citeauthoryear{Barbu, Mayo, Alverio, Luo, Wang, Gutfreund,
  Tenenbaum, and Katz}{Barbu et~al\mbox{.}}{2019}]%
        {Barbu2019-gw}
\bibfield{author}{\bibinfo{person}{Andrei Barbu}, \bibinfo{person}{David Mayo},
  \bibinfo{person}{Julian Alverio}, \bibinfo{person}{William Luo},
  \bibinfo{person}{Christopher Wang}, \bibinfo{person}{Dan Gutfreund},
  \bibinfo{person}{Josh Tenenbaum}, {and} \bibinfo{person}{Boris Katz}.}
  \bibinfo{year}{2019}\natexlab{}.
\newblock \showarticletitle{{ObjectNet}: A large-scale bias-controlled dataset
  for pushing the limits of object recognition models}. In
  \bibinfo{booktitle}{\emph{Advances in Neural Information Processing
  Systems}}, \bibfield{editor}{\bibinfo{person}{H~Wallach},
  \bibinfo{person}{H~Larochelle}, \bibinfo{person}{A~Beygelzimer},
  \bibinfo{person}{F~d\textbackslashtextquotesingle Alch{\'e}-Buc},
  \bibinfo{person}{E~Fox}, {and} \bibinfo{person}{R~Garnett}} (Eds.),
  Vol.~\bibinfo{volume}{32}. \bibinfo{publisher}{Curran Associates, Inc.}
\newblock
\urldef\tempurl%
\url{https://proceedings.neurips.cc/paper/2019/file/97af07a14cacba681feacf3012730892-Paper.pdf}
\showURL{%
\tempurl}


\bibitem[\protect\citeauthoryear{Belkin and Niyogi}{Belkin and Niyogi}{2006}]%
        {Belkin2006-tx}
\bibfield{author}{\bibinfo{person}{Mikhail Belkin} {and}
  \bibinfo{person}{Partha Niyogi}.} \bibinfo{year}{2006}\natexlab{}.
\newblock \bibinfo{title}{Manifold regularization: A geometric framework for
  learning from labeled and unlabeled examples}.
\newblock
  \bibinfo{howpublished}{\url{https://www.jmlr.org/papers/volume7/belkin06a/belkin06a.pdf}}.
\newblock
\urldef\tempurl%
\url{https://www.jmlr.org/papers/volume7/belkin06a/belkin06a.pdf}
\showURL{%
\tempurl}
\newblock
\shownote{Accessed: 2023-3-7.}


\bibitem[\protect\citeauthoryear{Bernhardsson}{Bernhardsson}{[n.d.]}]%
        {annoy}
\bibfield{author}{\bibinfo{person}{E. Bernhardsson}.}
  \bibinfo{year}{[n.d.]}\natexlab{}.
\newblock \bibinfo{title}{ANNOY: Approximate Nearest Neighbors Oh Yeah}.
\newblock \bibinfo{howpublished}{\url{https://github.com/spotify/annoy}}.
\newblock
\newblock
\shownote{Accessed: 2021-05-20.}


\bibitem[\protect\citeauthoryear{Burt and Adelson}{Burt and Adelson}{1983}]%
        {Burt1983-er}
\bibfield{author}{\bibinfo{person}{P Burt} {and} \bibinfo{person}{E Adelson}.}
  \bibinfo{year}{1983}\natexlab{}.
\newblock \showarticletitle{The Laplacian Pyramid as a Compact Image Code}.
\newblock \bibinfo{journal}{\emph{IEEE Trans. Commun.}} \bibinfo{volume}{31},
  \bibinfo{number}{4} (\bibinfo{date}{April} \bibinfo{year}{1983}),
  \bibinfo{pages}{532--540}.
\newblock
\showISSN{1558-0857}
\urldef\tempurl%
\url{https://doi.org/10.1109/TCOM.1983.1095851}
\showDOI{\tempurl}


\bibitem[\protect\citeauthoryear{Chen, Li, Yu, El~Kholy, Ahmed, Gan, Cheng, and
  Liu}{Chen et~al\mbox{.}}{2019}]%
        {Chen2019-iv}
\bibfield{author}{\bibinfo{person}{Yen-Chun Chen}, \bibinfo{person}{Linjie Li},
  \bibinfo{person}{Licheng Yu}, \bibinfo{person}{Ahmed El~Kholy},
  \bibinfo{person}{Faisal Ahmed}, \bibinfo{person}{Zhe Gan},
  \bibinfo{person}{Yu Cheng}, {and} \bibinfo{person}{Jingjing Liu}.}
  \bibinfo{year}{2019}\natexlab{}.
\newblock \showarticletitle{{UNITER}: {UNiversal} {Image-TExt} Representation
  Learning}.
\newblock  (\bibinfo{date}{Sept.} \bibinfo{year}{2019}).
\newblock
\showeprint[arxiv]{1909.11740}~[cs.CV]
\urldef\tempurl%
\url{http://arxiv.org/abs/1909.11740}
\showURL{%
\tempurl}


\bibitem[\protect\citeauthoryear{Cheney, Chiticariu, and Tan}{Cheney
  et~al\mbox{.}}{2009}]%
        {CheneyCT09}
\bibfield{author}{\bibinfo{person}{James Cheney}, \bibinfo{person}{Laura
  Chiticariu}, {and} \bibinfo{person}{Wang~Chiew Tan}.}
  \bibinfo{year}{2009}\natexlab{}.
\newblock \showarticletitle{Provenance in Databases: Why, How, and Where}.
\newblock \bibinfo{journal}{\emph{Found. Trends Databases}}
  \bibinfo{volume}{1}, \bibinfo{number}{4} (\bibinfo{year}{2009}),
  \bibinfo{pages}{379--474}.
\newblock
\urldef\tempurl%
\url{https://doi.org/10.1561/1900000006}
\showDOI{\tempurl}


\bibitem[\protect\citeauthoryear{Coleman, Chou, Katz-Samuels, and
  {others}}{Coleman et~al\mbox{.}}{2022}]%
        {Coleman2022-ag}
\bibfield{author}{\bibinfo{person}{C Coleman}, \bibinfo{person}{E Chou},
  \bibinfo{person}{J Katz-Samuels}, {and} \bibinfo{person}{{others}}.}
  \bibinfo{year}{2022}\natexlab{}.
\newblock \showarticletitle{Similarity search for efficient active learning and
  search of rare concepts}.
\newblock \bibinfo{journal}{\emph{Proceedings of the}} (\bibinfo{year}{2022}).
\newblock
\urldef\tempurl%
\url{https://ojs.aaai.org/index.php/AAAI/article/view/20591}
\showURL{%
\tempurl}


\bibitem[\protect\citeauthoryear{Cox, Miller, Minka, Papathomas, and
  Yianilos}{Cox et~al\mbox{.}}{2000}]%
        {Cox2000-ai}
\bibfield{author}{\bibinfo{person}{I~J Cox}, \bibinfo{person}{M~L Miller},
  \bibinfo{person}{T~P Minka}, \bibinfo{person}{T~V Papathomas}, {and}
  \bibinfo{person}{P~N Yianilos}.} \bibinfo{year}{2000}\natexlab{}.
\newblock \showarticletitle{The Bayesian image retrieval system, {PicHunter}:
  theory, implementation, and psychophysical experiments}.
\newblock \bibinfo{journal}{\emph{IEEE Trans. Image Process.}}
  \bibinfo{volume}{9}, \bibinfo{number}{1} (\bibinfo{date}{Jan.}
  \bibinfo{year}{2000}), \bibinfo{pages}{20--37}.
\newblock
\showISSN{1057-7149, 1941-0042}
\urldef\tempurl%
\url{https://doi.org/10.1109/83.817596}
\showDOI{\tempurl}


\bibitem[\protect\citeauthoryear{Doll{\'a}r, Appel, Belongie, and
  Perona}{Doll{\'a}r et~al\mbox{.}}{2014}]%
        {Dollar2014-nt}
\bibfield{author}{\bibinfo{person}{Piotr Doll{\'a}r}, \bibinfo{person}{Ron
  Appel}, \bibinfo{person}{Serge Belongie}, {and} \bibinfo{person}{Pietro
  Perona}.} \bibinfo{year}{2014}\natexlab{}.
\newblock \showarticletitle{Fast feature pyramids for object detection}.
\newblock \bibinfo{journal}{\emph{IEEE Trans. Pattern Anal. Mach. Intell.}}
  \bibinfo{volume}{36}, \bibinfo{number}{8} (\bibinfo{date}{Aug.}
  \bibinfo{year}{2014}), \bibinfo{pages}{1532--1545}.
\newblock
\showISSN{0162-8828, 1939-3539}
\urldef\tempurl%
\url{https://doi.org/10.1109/TPAMI.2014.2300479}
\showDOI{\tempurl}


\bibitem[\protect\citeauthoryear{Dong, Moses, and Li}{Dong
  et~al\mbox{.}}{2011}]%
        {Dong2011-hs}
\bibfield{author}{\bibinfo{person}{Wei Dong}, \bibinfo{person}{Charikar Moses},
  {and} \bibinfo{person}{Kai Li}.} \bibinfo{year}{2011}\natexlab{}.
\newblock \showarticletitle{Efficient k-nearest neighbor graph construction for
  generic similarity measures}. In \bibinfo{booktitle}{\emph{Proceedings of the
  20th international conference on World wide web - {WWW} '11}} (Hyderabad,
  India). \bibinfo{publisher}{ACM Press}, \bibinfo{address}{New York, New York,
  USA}.
\newblock
\showISBNx{9781450306324}
\urldef\tempurl%
\url{https://doi.org/10.1145/1963405.1963487}
\showDOI{\tempurl}


\bibitem[\protect\citeauthoryear{Elbassuoni, Ramanath, and Weikum}{Elbassuoni
  et~al\mbox{.}}{2011}]%
        {Elbassuoni2011-cw}
\bibfield{author}{\bibinfo{person}{Shady Elbassuoni}, \bibinfo{person}{Maya
  Ramanath}, {and} \bibinfo{person}{Gerhard Weikum}.}
  \bibinfo{year}{2011}\natexlab{}.
\newblock \showarticletitle{Query relaxation for entity-relationship search}.
\newblock In \bibinfo{booktitle}{\emph{The Semanic Web: Research and
  Applications}}. \bibinfo{publisher}{Springer Berlin Heidelberg},
  \bibinfo{address}{Berlin, Heidelberg}, \bibinfo{pages}{62--76}.
\newblock
\showISBNx{9783642210631, 9783642210648}
\showISSN{0302-9743, 1611-3349}
\urldef\tempurl%
\url{https://doi.org/10.1007/978-3-642-21064-8\_5}
\showDOI{\tempurl}


\bibitem[\protect\citeauthoryear{Faghri, Fleet, Kiros, and Fidler}{Faghri
  et~al\mbox{.}}{2017}]%
        {Faghri2017-kr}
\bibfield{author}{\bibinfo{person}{Fartash Faghri}, \bibinfo{person}{David~J
  Fleet}, \bibinfo{person}{Jamie~Ryan Kiros}, {and} \bibinfo{person}{Sanja
  Fidler}.} \bibinfo{year}{2017}\natexlab{}.
\newblock \showarticletitle{{VSE++}: Improving {Visual-Semantic} Embeddings
  with Hard Negatives}.
\newblock  (\bibinfo{date}{July} \bibinfo{year}{2017}).
\newblock
\showeprint[arxiv]{1707.05612}~[cs.LG]
\urldef\tempurl%
\url{http://arxiv.org/abs/1707.05612}
\showURL{%
\tempurl}


\bibitem[\protect\citeauthoryear{Frome, Corrado, Shlens, Bengio, Dean, Ranzato,
  and Mikolov}{Frome et~al\mbox{.}}{2013}]%
        {Frome2013-ue}
\bibfield{author}{\bibinfo{person}{Andrea Frome}, \bibinfo{person}{Greg~S
  Corrado}, \bibinfo{person}{Jon Shlens}, \bibinfo{person}{Samy Bengio},
  \bibinfo{person}{Jeff Dean}, \bibinfo{person}{Marc~Aurelio Ranzato}, {and}
  \bibinfo{person}{Tomas Mikolov}.} \bibinfo{year}{2013}\natexlab{}.
\newblock \showarticletitle{{DeViSE}: A Deep {Visual-Semantic} Embedding
  Model}. In \bibinfo{booktitle}{\emph{Advances in Neural Information
  Processing Systems}}, \bibfield{editor}{\bibinfo{person}{C~J Burges},
  \bibinfo{person}{L~Bottou}, \bibinfo{person}{M~Welling},
  \bibinfo{person}{Z~Ghahramani}, {and} \bibinfo{person}{K~Q Weinberger}}
  (Eds.), Vol.~\bibinfo{volume}{26}. \bibinfo{publisher}{Curran Associates,
  Inc.}
\newblock
\urldef\tempurl%
\url{https://proceedings.neurips.cc/paper/2013/file/7cce53cf90577442771720a370c3c723-Paper.pdf}
\showURL{%
\tempurl}


\bibitem[\protect\citeauthoryear{Garnett, Krishnamurthy, Xiong, Schneider, and
  Mann}{Garnett et~al\mbox{.}}{2012a}]%
        {Garnett2012-tp}
\bibfield{author}{\bibinfo{person}{Roman Garnett}, \bibinfo{person}{Yamuna
  Krishnamurthy}, \bibinfo{person}{Xuehan Xiong}, \bibinfo{person}{Jeff
  Schneider}, {and} \bibinfo{person}{Richard Mann}.}
  \bibinfo{year}{2012}\natexlab{a}.
\newblock \showarticletitle{Bayesian Optimal Active Search and Surveying}.
\newblock  (\bibinfo{date}{June} \bibinfo{year}{2012}).
\newblock
\showeprint[arxiv]{1206.6406}~[cs.LG]
\urldef\tempurl%
\url{http://arxiv.org/abs/1206.6406}
\showURL{%
\tempurl}


\bibitem[\protect\citeauthoryear{Garnett, Krishnamurthy, Xiong, Schneider, and
  Mann}{Garnett et~al\mbox{.}}{2012b}]%
        {garnett_active_search}
\bibfield{author}{\bibinfo{person}{Roman Garnett}, \bibinfo{person}{Yamuna
  Krishnamurthy}, \bibinfo{person}{Xuehan Xiong}, \bibinfo{person}{Jeff
  Schneider}, {and} \bibinfo{person}{Richard Mann}.}
  \bibinfo{year}{2012}\natexlab{b}.
\newblock \showarticletitle{Bayesian Optimal Active Search and Surveying}.
\newblock  (\bibinfo{date}{June} \bibinfo{year}{2012}).
\newblock
\showeprint[arxiv]{1206.6406}~[cs.LG]


\bibitem[\protect\citeauthoryear{Grossman, Cormack, and Roegiest}{Grossman
  et~al\mbox{.}}{2016}]%
        {Grossman2016-total_recall_trec}
\bibfield{author}{\bibinfo{person}{Maura~R Grossman}, \bibinfo{person}{G
  Cormack}, {and} \bibinfo{person}{Adam Roegiest}.}
  \bibinfo{year}{2016}\natexlab{}.
\newblock \showarticletitle{{TREC} 2016 Total Recall Track Overview}.
\newblock \bibinfo{journal}{\emph{TREC}} (\bibinfo{year}{2016}).
\newblock


\bibitem[\protect\citeauthoryear{Gupta, Dollár, and Girshick}{Gupta
  et~al\mbox{.}}{2019}]%
        {lvis_dataset}
\bibfield{author}{\bibinfo{person}{Agrim Gupta}, \bibinfo{person}{Piotr
  Dollár}, {and} \bibinfo{person}{Ross Girshick}.}
  \bibinfo{year}{2019}\natexlab{}.
\newblock \showarticletitle{LVIS: A Dataset for Large Vocabulary Instance
  Segmentation}.
\newblock \bibinfo{journal}{\emph{arXiv [cs.CV]}} (\bibinfo{date}{Aug}
  \bibinfo{year}{2019}).
\newblock
\urldef\tempurl%
\url{https://arxiv.org/abs/1908.03195}
\showURL{%
\tempurl}


\bibitem[\protect\citeauthoryear{Insights}{Insights}{[n.d.]}]%
        {wildlife_insights}
\bibfield{author}{\bibinfo{person}{Wildlife Insights}.}
  \bibinfo{year}{[n.d.]}\natexlab{}.
\newblock \bibinfo{title}{Wildlife Insights}.
\newblock
\newblock
\urldef\tempurl%
\url{https://www.wildlifeinsights.org/}
\showURL{%
\tempurl}
\newblock
\shownote{Accessed on Mar 26, 2023.}


\bibitem[\protect\citeauthoryear{Ishikawa, Subramanya, and Faloutsos}{Ishikawa
  et~al\mbox{.}}{1998}]%
        {Ishikawa1998-dx}
\bibfield{author}{\bibinfo{person}{Y Ishikawa}, \bibinfo{person}{R Subramanya},
  {and} \bibinfo{person}{C Faloutsos}.} \bibinfo{year}{1998}\natexlab{}.
\newblock \showarticletitle{{MindReader}: Querying Databases Through Multiple
  Examples}.
\newblock \bibinfo{journal}{\emph{VLDB J.}} (\bibinfo{year}{1998}).
\newblock
\showISSN{1066-8888}
\urldef\tempurl%
\url{https://www.semanticscholar.org/paper/04938be9fd727ea6363cc950efd263ff82d02b77}
\showURL{%
\tempurl}


\bibitem[\protect\citeauthoryear{Jia, Yang, Xia, Chen, Parekh, Pham, Le, Sung,
  Li, and Duerig}{Jia et~al\mbox{.}}{2021}]%
        {Jia2021-np}
\bibfield{author}{\bibinfo{person}{Chao Jia}, \bibinfo{person}{Yinfei Yang},
  \bibinfo{person}{Ye Xia}, \bibinfo{person}{Yi-Ting Chen},
  \bibinfo{person}{Zarana Parekh}, \bibinfo{person}{Hieu Pham},
  \bibinfo{person}{Quoc~V Le}, \bibinfo{person}{Yunhsuan Sung},
  \bibinfo{person}{Zhen Li}, {and} \bibinfo{person}{Tom Duerig}.}
  \bibinfo{year}{2021}\natexlab{}.
\newblock \showarticletitle{Scaling up visual and vision-language
  representation learning with noisy text supervision}.
\newblock  (\bibinfo{date}{Feb.} \bibinfo{year}{2021}).
\newblock
\showeprint[arxiv]{2102.05918}~[cs.CV]
\urldef\tempurl%
\url{http://proceedings.mlr.press/v139/jia21b/jia21b.pdf}
\showURL{%
\tempurl}


\bibitem[\protect\citeauthoryear{Jiang, Malkomes, Converse, Shofner, Moseley,
  and Garnett}{Jiang et~al\mbox{.}}{2017}]%
        {Jiang2017-iy}
\bibfield{author}{\bibinfo{person}{Shali Jiang}, \bibinfo{person}{Gustavo
  Malkomes}, \bibinfo{person}{Geoff Converse}, \bibinfo{person}{Alyssa
  Shofner}, \bibinfo{person}{Benjamin Moseley}, {and} \bibinfo{person}{Roman
  Garnett}.} \bibinfo{year}{2017}\natexlab{}.
\newblock \showarticletitle{Efficient Nonmyopic Active Search}. In
  \bibinfo{booktitle}{\emph{Proceedings of the 34th International Conference on
  Machine Learning}} \emph{(\bibinfo{series}{Proceedings of Machine Learning
  Research}, Vol.~\bibinfo{volume}{70})},
  \bibfield{editor}{\bibinfo{person}{Doina Precup} {and}
  \bibinfo{person}{Yee~Whye Teh}} (Eds.). \bibinfo{publisher}{PMLR},
  \bibinfo{pages}{1714--1723}.
\newblock
\urldef\tempurl%
\url{https://proceedings.mlr.press/v70/jiang17d.html}
\showURL{%
\tempurl}


\bibitem[\protect\citeauthoryear{Jing, Li, Zhang, and Zhang}{Jing
  et~al\mbox{.}}{2004}]%
        {Jing2004-ex}
\bibfield{author}{\bibinfo{person}{F Jing}, \bibinfo{person}{M Li},
  \bibinfo{person}{H-J Zhang}, {and} \bibinfo{person}{B Zhang}.}
  \bibinfo{year}{2004}\natexlab{}.
\newblock \showarticletitle{Relevance feedback in region-based image
  retrieval}.
\newblock \bibinfo{journal}{\emph{IEEE Trans. Circuits Syst. Video Technol.}}
  \bibinfo{volume}{14}, \bibinfo{number}{5} (\bibinfo{date}{May}
  \bibinfo{year}{2004}), \bibinfo{pages}{672--681}.
\newblock
\showISSN{1051-8215, 1558-2205}
\urldef\tempurl%
\url{https://doi.org/10.1109/tcsvt.2004.826775}
\showDOI{\tempurl}


\bibitem[\protect\citeauthoryear{Kasai, Qian, Gurajada, Li, and Popa}{Kasai
  et~al\mbox{.}}{2019}]%
        {Kasai2019-pu}
\bibfield{author}{\bibinfo{person}{Jungo Kasai}, \bibinfo{person}{Kun Qian},
  \bibinfo{person}{Sairam Gurajada}, \bibinfo{person}{Yunyao Li}, {and}
  \bibinfo{person}{Lucian Popa}.} \bibinfo{year}{2019}\natexlab{}.
\newblock \showarticletitle{Low-resource deep entity resolution with transfer
  and active learning}. In \bibinfo{booktitle}{\emph{Proceedings of the 57th
  Annual Meeting of the Association for Computational Linguistics}} (Florence,
  Italy). \bibinfo{publisher}{Association for Computational Linguistics},
  \bibinfo{address}{Stroudsburg, PA, USA}.
\newblock
\urldef\tempurl%
\url{https://doi.org/10.18653/v1/p19-1586}
\showDOI{\tempurl}


\bibitem[\protect\citeauthoryear{Kouw and Loog}{Kouw and Loog}{2021}]%
        {Kouw2021-vr}
\bibfield{author}{\bibinfo{person}{Wouter~M Kouw} {and} \bibinfo{person}{Marco
  Loog}.} \bibinfo{year}{2021}\natexlab{}.
\newblock \showarticletitle{A review of domain adaptation without target
  labels}.
\newblock \bibinfo{journal}{\emph{IEEE Trans. Pattern Anal. Mach. Intell.}}
  \bibinfo{volume}{43}, \bibinfo{number}{3} (\bibinfo{date}{March}
  \bibinfo{year}{2021}), \bibinfo{pages}{766--785}.
\newblock
\showISSN{0162-8828, 1939-3539}
\urldef\tempurl%
\url{https://doi.org/10.1109/TPAMI.2019.2945942}
\showDOI{\tempurl}


\bibitem[\protect\citeauthoryear{Krishnan, Wang, Wu, Franklin, and
  Goldberg}{Krishnan et~al\mbox{.}}{2016}]%
        {Krishnan2016-aa}
\bibfield{author}{\bibinfo{person}{Sanjay Krishnan}, \bibinfo{person}{Jiannan
  Wang}, \bibinfo{person}{Eugene Wu}, \bibinfo{person}{Michael~J Franklin},
  {and} \bibinfo{person}{Ken Goldberg}.} \bibinfo{year}{2016}\natexlab{}.
\newblock \showarticletitle{{ActiveClean}}.
\newblock \bibinfo{journal}{\emph{Proceedings VLDB Endowment}}
  \bibinfo{volume}{9}, \bibinfo{number}{12} (\bibinfo{date}{Aug.}
  \bibinfo{year}{2016}), \bibinfo{pages}{948--959}.
\newblock
\showISSN{2150-8097}
\urldef\tempurl%
\url{https://doi.org/10.14778/2994509.2994514}
\showDOI{\tempurl}


\bibitem[\protect\citeauthoryear{Lin, Maire, Belongie, Bourdev, Girshick, Hays,
  Perona, Ramanan, Doll{\'{a}}r, and Zitnick}{Lin et~al\mbox{.}}{2014}]%
        {coco}
\bibfield{author}{\bibinfo{person}{Tsung{-}Yi Lin}, \bibinfo{person}{Michael
  Maire}, \bibinfo{person}{Serge~J. Belongie}, \bibinfo{person}{Lubomir~D.
  Bourdev}, \bibinfo{person}{Ross~B. Girshick}, \bibinfo{person}{James Hays},
  \bibinfo{person}{Pietro Perona}, \bibinfo{person}{Deva Ramanan},
  \bibinfo{person}{Piotr Doll{\'{a}}r}, {and} \bibinfo{person}{C.~Lawrence
  Zitnick}.} \bibinfo{year}{2014}\natexlab{}.
\newblock \showarticletitle{Microsoft {COCO:} Common Objects in Context}.
\newblock \bibinfo{journal}{\emph{CoRR}}  \bibinfo{volume}{abs/1405.0312}
  (\bibinfo{year}{2014}).
\newblock
\showeprint[arxiv]{1405.0312}
\urldef\tempurl%
\url{http://arxiv.org/abs/1405.0312}
\showURL{%
\tempurl}


\bibitem[\protect\citeauthoryear{Lin, Doll{\'a}r, Girshick, He, Hariharan, and
  Belongie}{Lin et~al\mbox{.}}{2016}]%
        {Lin2016-vr}
\bibfield{author}{\bibinfo{person}{Tsung-Yi Lin}, \bibinfo{person}{Piotr
  Doll{\'a}r}, \bibinfo{person}{Ross Girshick}, \bibinfo{person}{Kaiming He},
  \bibinfo{person}{Bharath Hariharan}, {and} \bibinfo{person}{Serge Belongie}.}
  \bibinfo{year}{2016}\natexlab{}.
\newblock \showarticletitle{Feature Pyramid Networks for Object Detection}.
\newblock  (\bibinfo{date}{Dec.} \bibinfo{year}{2016}).
\newblock
\showeprint[arxiv]{1612.03144}~[cs.CV]
\urldef\tempurl%
\url{http://arxiv.org/abs/1612.03144}
\showURL{%
\tempurl}


\bibitem[\protect\citeauthoryear{Liu and Nocedal}{Liu and Nocedal}{1989}]%
        {Liu1989-mz}
\bibfield{author}{\bibinfo{person}{Dong~C Liu} {and} \bibinfo{person}{Jorge
  Nocedal}.} \bibinfo{year}{1989}\natexlab{}.
\newblock \showarticletitle{On the limited memory {BFGS} method for large scale
  optimization}.
\newblock \bibinfo{journal}{\emph{Math. Program.}} \bibinfo{volume}{45},
  \bibinfo{number}{1} (\bibinfo{date}{Aug.} \bibinfo{year}{1989}),
  \bibinfo{pages}{503--528}.
\newblock
\showISSN{0025-5610, 1436-4646}
\urldef\tempurl%
\url{https://doi.org/10.1007/BF01589116}
\showDOI{\tempurl}


\bibitem[\protect\citeauthoryear{Manning, Raghavan, and Sch{\"u}tze}{Manning
  et~al\mbox{.}}{2008a}]%
        {irbook}
\bibfield{author}{\bibinfo{person}{C.D. Manning}, \bibinfo{person}{P.
  Raghavan}, {and} \bibinfo{person}{H. Sch{\"u}tze}.}
  \bibinfo{year}{2008}\natexlab{a}.
\newblock \bibinfo{booktitle}{\emph{Introduction to Information Retrieval}}.
\newblock \bibinfo{publisher}{Cambridge University Press}.
\newblock
\showISBNx{9781139472104}
\urldef\tempurl%
\url{https://books.google.com/books?id=t1PoSh4uwVcC}
\showURL{%
\tempurl}


\bibitem[\protect\citeauthoryear{Manning, Raghavan, and Sch{\"u}tze}{Manning
  et~al\mbox{.}}{2008b}]%
        {Manning2008-ua}
\bibfield{author}{\bibinfo{person}{Christopher~D Manning},
  \bibinfo{person}{Prabhakar Raghavan}, {and} \bibinfo{person}{Hinrich
  Sch{\"u}tze}.} \bibinfo{year}{2008}\natexlab{b}.
\newblock \bibinfo{booktitle}{\emph{Introduction to Information Retrieval}}.
\newblock \bibinfo{publisher}{Cambridge University Press}.
\newblock
\showISBNx{9780511809071}
\urldef\tempurl%
\url{https://doi.org/10.1017/CBO9780511809071}
\showDOI{\tempurl}


\bibitem[\protect\citeauthoryear{Murphy}{Murphy}{2022}]%
        {Murphy2022-au}
\bibfield{author}{\bibinfo{person}{Kevin~P Murphy}.}
  \bibinfo{year}{2022}\natexlab{}.
\newblock \bibinfo{booktitle}{\emph{Probabilistic Machine Learning: An
  Introduction}}.
\newblock \bibinfo{publisher}{MIT Press}.
\newblock
\showISBNx{9780262046824}
\urldef\tempurl%
\url{https://play.google.com/store/books/details?id=wrZNEAAAQBAJ}
\showURL{%
\tempurl}


\bibitem[\protect\citeauthoryear{Niculescu-Mizil and Caruana}{Niculescu-Mizil
  and Caruana}{2005}]%
        {Niculescu-Mizil2005-hg}
\bibfield{author}{\bibinfo{person}{Alexandru Niculescu-Mizil} {and}
  \bibinfo{person}{Rich Caruana}.} \bibinfo{year}{2005}\natexlab{}.
\newblock \showarticletitle{Predicting good probabilities with supervised
  learning}. In \bibinfo{booktitle}{\emph{Proceedings of the 22nd international
  conference on Machine learning - {ICML} '05}} (Bonn, Germany).
  \bibinfo{publisher}{ACM Press}, \bibinfo{address}{New York, New York, USA}.
\newblock
\showISBNx{9781595931801}
\urldef\tempurl%
\url{https://doi.org/10.1145/1102351.1102430}
\showDOI{\tempurl}


\bibitem[\protect\citeauthoryear{Paszke, Gross, Massa, Lerer, Bradbury, Chanan,
  Killeen, Lin, Gimelshein, Antiga, Desmaison, K{\"o}pf, Yang, DeVito, Raison,
  Tejani, Chilamkurthy, Steiner, Fang, Bai, and Chintala}{Paszke
  et~al\mbox{.}}{2019}]%
        {Paszke2019-mp}
\bibfield{author}{\bibinfo{person}{Adam Paszke}, \bibinfo{person}{Sam Gross},
  \bibinfo{person}{Francisco Massa}, \bibinfo{person}{Adam Lerer},
  \bibinfo{person}{James Bradbury}, \bibinfo{person}{Gregory Chanan},
  \bibinfo{person}{Trevor Killeen}, \bibinfo{person}{Zeming Lin},
  \bibinfo{person}{Natalia Gimelshein}, \bibinfo{person}{Luca Antiga},
  \bibinfo{person}{Alban Desmaison}, \bibinfo{person}{Andreas K{\"o}pf},
  \bibinfo{person}{Edward Yang}, \bibinfo{person}{Zach DeVito},
  \bibinfo{person}{Martin Raison}, \bibinfo{person}{Alykhan Tejani},
  \bibinfo{person}{Sasank Chilamkurthy}, \bibinfo{person}{Benoit Steiner},
  \bibinfo{person}{Lu Fang}, \bibinfo{person}{Junjie Bai}, {and}
  \bibinfo{person}{Soumith Chintala}.} \bibinfo{year}{2019}\natexlab{}.
\newblock \showarticletitle{{PyTorch}: An Imperative Style, {High-Performance}
  Deep Learning Library}.
\newblock  (\bibinfo{date}{Dec.} \bibinfo{year}{2019}).
\newblock
\showeprint[arxiv]{1912.01703}~[cs.LG]
\urldef\tempurl%
\url{http://arxiv.org/abs/1912.01703}
\showURL{%
\tempurl}


\bibitem[\protect\citeauthoryear{Platt}{Platt}{2000}]%
        {Platt2000-tc}
\bibfield{author}{\bibinfo{person}{John~C Platt}.}
  \bibinfo{year}{2000}\natexlab{}.
\newblock \showarticletitle{Probabilistic Outputs for Support Vector Machines
  and Comparisons to Regularized Likelihood Methods}.
\newblock  \bibinfo{volume}{10}, \bibinfo{number}{3} (\bibinfo{date}{June}
  \bibinfo{year}{2000}).
\newblock
\urldef\tempurl%
\url{http://dx.doi.org/}
\showURL{%
\tempurl}


\bibitem[\protect\citeauthoryear{Radford, Kim, Hallacy, Ramesh, Goh, Agarwal,
  Sastry, Askell, Mishkin, Clark, and et~al.}{Radford et~al\mbox{.}}{[n.d.]}]%
        {clip_model}
\bibfield{author}{\bibinfo{person}{Alec Radford}, \bibinfo{person}{1.~Jong~Wook
  Kim}, \bibinfo{person}{1.~Chris Hallacy}, \bibinfo{person}{Aditya Ramesh},
  \bibinfo{person}{Gabriel Goh}, \bibinfo{person}{Sandhini Agarwal},
  \bibinfo{person}{Girish Sastry}, \bibinfo{person}{Amanda Askell},
  \bibinfo{person}{Pamela Mishkin}, \bibinfo{person}{Jack Clark}, {and}
  \bibinfo{person}{et al.}} \bibinfo{year}{[n.d.]}\natexlab{}.
\newblock \bibinfo{title}{Learning transferable visual models from natural
  language supervision}.
\newblock
\newblock
\urldef\tempurl%
\url{https://cdn.openai.com/papers/Learning_Transferable_Visual_Models_From_Natural_Language_Supervision.pdf}
\showURL{%
\tempurl}


\bibitem[\protect\citeauthoryear{Radford, Kim, Hallacy, Ramesh, Goh, Agarwal,
  Sastry, Askell, Mishkin, Clark, Krueger, and Sutskever}{Radford
  et~al\mbox{.}}{2021}]%
        {Radford2021-rk}
\bibfield{author}{\bibinfo{person}{Alec Radford}, \bibinfo{person}{Jong~Wook
  Kim}, \bibinfo{person}{Chris Hallacy}, \bibinfo{person}{Aditya Ramesh},
  \bibinfo{person}{Gabriel Goh}, \bibinfo{person}{Sandhini Agarwal},
  \bibinfo{person}{Girish Sastry}, \bibinfo{person}{Amanda Askell},
  \bibinfo{person}{Pamela Mishkin}, \bibinfo{person}{Jack Clark},
  \bibinfo{person}{Gretchen Krueger}, {and} \bibinfo{person}{Ilya Sutskever}.}
  \bibinfo{year}{2021}\natexlab{}.
\newblock \showarticletitle{Learning Transferable Visual Models From Natural
  Language Supervision}.
\newblock  (\bibinfo{date}{Feb.} \bibinfo{year}{2021}).
\newblock
\showeprint[arxiv]{2103.00020}~[cs.CV]
\urldef\tempurl%
\url{http://arxiv.org/abs/2103.00020}
\showURL{%
\tempurl}


\bibitem[\protect\citeauthoryear{Renders}{Renders}{2018}]%
        {renders_2017}
\bibfield{author}{\bibinfo{person}{Jean-Michel Renders}.}
  \bibinfo{year}{2018}\natexlab{}.
\newblock \showarticletitle{Active search for high recall: A non-stationary
  extension of Thompson sampling}.
\newblock In \bibinfo{booktitle}{\emph{Lecture Notes in Computer Science}}.
  \bibinfo{publisher}{Springer International Publishing},
  \bibinfo{address}{Cham}, \bibinfo{pages}{722--728}.
\newblock


\bibitem[\protect\citeauthoryear{Rocchio}{Rocchio}{1971}]%
        {rocchio71}
\bibfield{author}{\bibinfo{person}{J.~J. Rocchio}.}
  \bibinfo{year}{1971}\natexlab{}.
\newblock \showarticletitle{Relevance feedback in information retrieval}.
\newblock In \bibinfo{booktitle}{\emph{The {SMART} Retrieval System --
  Experiments in Automatic Document Processing}},
  \bibfield{editor}{\bibinfo{person}{Gerard Salton}} (Ed.).
  \bibinfo{publisher}{Prentice Hall}, \bibinfo{address}{Englewood Cliffs, NJ},
  \bibinfo{pages}{313--323}.
\newblock


\bibitem[\protect\citeauthoryear{Salton and Buckley}{Salton and
  Buckley}{1990}]%
        {Salton1990-ks}
\bibfield{author}{\bibinfo{person}{Gerard Salton} {and} \bibinfo{person}{Chris
  Buckley}.} \bibinfo{year}{1990}\natexlab{}.
\newblock \showarticletitle{Improving retrieval performance by relevance
  feedback}.
\newblock \bibinfo{journal}{\emph{J. Am. Soc. Inf. Sci.}} \bibinfo{volume}{41},
  \bibinfo{number}{4} (\bibinfo{date}{June} \bibinfo{year}{1990}),
  \bibinfo{pages}{288--297}.
\newblock
\showISSN{0002-8231, 1097-4571}
\urldef\tempurl%
\url{https://doi.org/10.1002/(sici)1097-4571(199006)41:4<288::aid-asi8>3.0.co;2-h}
\showDOI{\tempurl}


\bibitem[\protect\citeauthoryear{Settles}{Settles}{[n.d.]}]%
        {Settles_undated-nl}
\bibfield{author}{\bibinfo{person}{Burr Settles}.}
  \bibinfo{year}{[n.d.]}\natexlab{}.
\newblock \bibinfo{booktitle}{\emph{Active Learning}}.
\newblock \bibinfo{publisher}{Morgan Claypool}.
\newblock
\showISBNx{9781608457267}


\bibitem[\protect\citeauthoryear{Shalev-Shwartz and Ben-David}{Shalev-Shwartz
  and Ben-David}{2014}]%
        {Shalev-Shwartz2014-kf}
\bibfield{author}{\bibinfo{person}{Shai Shalev-Shwartz} {and}
  \bibinfo{person}{Shai Ben-David}.} \bibinfo{year}{2014}\natexlab{}.
\newblock \bibinfo{booktitle}{\emph{Understanding Machine Learning: From Theory
  to Algorithms}}.
\newblock \bibinfo{publisher}{Cambridge University Press}.
\newblock
\showISBNx{9781107057135, 9781107298019}
\urldef\tempurl%
\url{https://doi.org/10.1017/CBO9781107298019}
\showDOI{\tempurl}


\bibitem[\protect\citeauthoryear{Simoncelli and Freeman}{Simoncelli and
  Freeman}{1995}]%
        {Simoncelli1995-bm}
\bibfield{author}{\bibinfo{person}{E~P Simoncelli} {and} \bibinfo{person}{W~T
  Freeman}.} \bibinfo{year}{1995}\natexlab{}.
\newblock \showarticletitle{The steerable pyramid: a flexible architecture for
  multi-scale derivative computation}. In
  \bibinfo{booktitle}{\emph{Proceedings., International Conference on Image
  Processing}}, Vol.~\bibinfo{volume}{3}. \bibinfo{pages}{444--447 vol.3}.
\newblock
\urldef\tempurl%
\url{https://doi.org/10.1109/ICIP.1995.537667}
\showDOI{\tempurl}


\bibitem[\protect\citeauthoryear{Tan, Cascante{-}Bonilla, Guo, Wu, Feng, and
  Ordonez}{Tan et~al\mbox{.}}{2019}]%
        {drilldown}
\bibfield{author}{\bibinfo{person}{Fuwen Tan}, \bibinfo{person}{Paola
  Cascante{-}Bonilla}, \bibinfo{person}{Xiaoxiao Guo}, \bibinfo{person}{Hui
  Wu}, \bibinfo{person}{Song Feng}, {and} \bibinfo{person}{Vicente Ordonez}.}
  \bibinfo{year}{2019}\natexlab{}.
\newblock \showarticletitle{Drill-down: Interactive Retrieval of Complex Scenes
  using Natural Language Queries}. In \bibinfo{booktitle}{\emph{Advances in
  Neural Information Processing Systems 32: Annual Conference on Neural
  Information Processing Systems 2019, NeurIPS 2019, December 8-14, 2019,
  Vancouver, BC, Canada}}, \bibfield{editor}{\bibinfo{person}{Hanna~M.
  Wallach}, \bibinfo{person}{Hugo Larochelle}, \bibinfo{person}{Alina
  Beygelzimer}, \bibinfo{person}{Florence d'Alch{\'{e}}{-}Buc},
  \bibinfo{person}{Emily~B. Fox}, {and} \bibinfo{person}{Roman Garnett}}
  (Eds.). \bibinfo{pages}{2647--2657}.
\newblock


\bibitem[\protect\citeauthoryear{Van~Horn, Branson, Farrell, Haber, Barry,
  Ipeirotis, Perona, and Belongie}{Van~Horn et~al\mbox{.}}{2015}]%
        {Van_Horn2015-zu}
\bibfield{author}{\bibinfo{person}{Grant Van~Horn}, \bibinfo{person}{Steve
  Branson}, \bibinfo{person}{Ryan Farrell}, \bibinfo{person}{Scott Haber},
  \bibinfo{person}{Jessie Barry}, \bibinfo{person}{Panos Ipeirotis},
  \bibinfo{person}{Pietro Perona}, {and} \bibinfo{person}{Serge Belongie}.}
  \bibinfo{year}{2015}\natexlab{}.
\newblock \showarticletitle{Building a bird recognition app and large scale
  dataset with citizen scientists: The fine print in fine-grained dataset
  collection}. In \bibinfo{booktitle}{\emph{2015 {IEEE} Conference on Computer
  Vision and Pattern Recognition ({CVPR})}} (Boston, MA, USA).
  \bibinfo{publisher}{IEEE}.
\newblock
\showISBNx{9781467369640}
\urldef\tempurl%
\url{https://doi.org/10.1109/cvpr.2015.7298658}
\showDOI{\tempurl}


\bibitem[\protect\citeauthoryear{Vasconcelos and Lippman}{Vasconcelos and
  Lippman}{[n.d.]}]%
        {Vasconcelos_undated-ja}
\bibfield{author}{\bibinfo{person}{Nuno Vasconcelos} {and}
  \bibinfo{person}{Andrew Lippman}.} \bibinfo{year}{[n.d.]}\natexlab{}.
\newblock \bibinfo{title}{Learning from user feedback in image retrieval
  systems}.
\newblock
  \bibinfo{howpublished}{\url{https://papers.nips.cc/paper/1999/file/7283518d47a05a09d33779a17adf1707-Paper.pdf}}.
\newblock
\urldef\tempurl%
\url{https://papers.nips.cc/paper/1999/file/7283518d47a05a09d33779a17adf1707-Paper.pdf}
\showURL{%
\tempurl}
\newblock
\shownote{Accessed: 2021-8-12.}


\bibitem[\protect\citeauthoryear{Wu, Faloutsos, Sycara, and Payne}{Wu
  et~al\mbox{.}}{[n.d.]}]%
        {Wu_undated-xy}
\bibfield{author}{\bibinfo{person}{Leejay Wu}, \bibinfo{person}{Christos
  Faloutsos}, \bibinfo{person}{Katia Sycara}, {and} \bibinfo{person}{Terry~R
  Payne}.} \bibinfo{year}{[n.d.]}\natexlab{}.
\newblock \bibinfo{title}{{FALCON}: Feedback adaptive loop for content-based
  retrieval}.
\newblock
  \bibinfo{howpublished}{\url{http://www.cs.cmu.edu/~christos/PUBLICATIONS/vldb2k-falcon.pdf}}.
\newblock
\urldef\tempurl%
\url{http://www.cs.cmu.edu/~christos/PUBLICATIONS/vldb2k-falcon.pdf}
\showURL{%
\tempurl}
\newblock
\shownote{Accessed: 2022-5-30.}


\bibitem[\protect\citeauthoryear{Yan, Zheng, Ives, Talukdar, and Yu}{Yan
  et~al\mbox{.}}{2013}]%
        {Yan2013-eh}
\bibfield{author}{\bibinfo{person}{Zhepeng Yan}, \bibinfo{person}{Nan Zheng},
  \bibinfo{person}{Zachary~G Ives}, \bibinfo{person}{Partha~Pratim Talukdar},
  {and} \bibinfo{person}{Cong Yu}.} \bibinfo{year}{2013}\natexlab{}.
\newblock \showarticletitle{Actively soliciting feedback for query answers in
  keyword search-based data integration}.
\newblock \bibinfo{journal}{\emph{Proceedings VLDB Endowment}}
  \bibinfo{volume}{6}, \bibinfo{number}{3} (\bibinfo{date}{Jan.}
  \bibinfo{year}{2013}), \bibinfo{pages}{205--216}.
\newblock
\showISSN{2150-8097}
\urldef\tempurl%
\url{https://doi.org/10.14778/2535569.2448954}
\showDOI{\tempurl}


\bibitem[\protect\citeauthoryear{Young, Rode-Margono, and Amin}{Young
  et~al\mbox{.}}{2018}]%
        {Young2018-xv}
\bibfield{author}{\bibinfo{person}{Stuart Young}, \bibinfo{person}{Johanna
  Rode-Margono}, {and} \bibinfo{person}{Rajan Amin}.}
  \bibinfo{year}{2018}\natexlab{}.
\newblock \showarticletitle{Software to facilitate and streamline camera trap
  data management: A review}.
\newblock \bibinfo{journal}{\emph{Ecol. Evol.}} \bibinfo{volume}{8},
  \bibinfo{number}{19} (\bibinfo{date}{Oct.} \bibinfo{year}{2018}),
  \bibinfo{pages}{9947--9957}.
\newblock
\showISSN{2045-7758}
\urldef\tempurl%
\url{https://doi.org/10.1002/ece3.4464}
\showDOI{\tempurl}


\bibitem[\protect\citeauthoryear{Yu, Xian, Chen, Liu, Liao, Madhavan, and
  Darrell}{Yu et~al\mbox{.}}{2018}]%
        {bdd}
\bibfield{author}{\bibinfo{person}{Fisher Yu}, \bibinfo{person}{Wenqi Xian},
  \bibinfo{person}{Yingying Chen}, \bibinfo{person}{Fangchen Liu},
  \bibinfo{person}{Mike Liao}, \bibinfo{person}{Vashisht Madhavan}, {and}
  \bibinfo{person}{Trevor Darrell}.} \bibinfo{year}{2018}\natexlab{}.
\newblock \showarticletitle{{BDD100K:} {A} Diverse Driving Video Database with
  Scalable Annotation Tooling}.
\newblock \bibinfo{journal}{\emph{CoRR}}  \bibinfo{volume}{abs/1805.04687}
  (\bibinfo{year}{2018}).
\newblock
\showeprint[arxiv]{1805.04687}
\urldef\tempurl%
\url{http://arxiv.org/abs/1805.04687}
\showURL{%
\tempurl}


\bibitem[\protect\citeauthoryear{Yu and Menzies}{Yu and Menzies}{2018}]%
        {Yu_menzies_2018}
\bibfield{author}{\bibinfo{person}{Zhe Yu} {and} \bibinfo{person}{Tim
  Menzies}.} \bibinfo{year}{2018}\natexlab{}.
\newblock \showarticletitle{Total Recall, Language Processing, and Software
  Engineering}.
\newblock  (\bibinfo{date}{Aug.} \bibinfo{year}{2018}).
\newblock
\showeprint[arxiv]{1809.00039}~[cs.SE]


\bibitem[\protect\citeauthoryear{Zhou and Huang}{Zhou and Huang}{2003}]%
        {Zhou2003-hb}
\bibfield{author}{\bibinfo{person}{Xiang~Sean Zhou} {and}
  \bibinfo{person}{Thomas~S Huang}.} \bibinfo{year}{2003}\natexlab{}.
\newblock \showarticletitle{Relevance feedback in image retrieval: A
  comprehensive review}.
\newblock \bibinfo{journal}{\emph{Multimed. Syst.}} \bibinfo{volume}{8},
  \bibinfo{number}{6} (\bibinfo{date}{April} \bibinfo{year}{2003}),
  \bibinfo{pages}{536--544}.
\newblock
\showISSN{0942-4962, 1432-1882}
\urldef\tempurl%
\url{https://doi.org/10.1007/s00530-002-0070-3}
\showDOI{\tempurl}


\bibitem[\protect\citeauthoryear{Zhu and Ghahramani}{Zhu and
  Ghahramani}{2002}]%
        {Zhu2002-hd}
\bibfield{author}{\bibinfo{person}{Xiaojin Zhu} {and} \bibinfo{person}{Zoubin
  Ghahramani}.} \bibinfo{year}{2002}\natexlab{}.
\newblock \showarticletitle{Learning from labeled and unlabeled data with label
  propagation}.
\newblock  (\bibinfo{year}{2002}).
\newblock
\urldef\tempurl%
\url{https://www.semanticscholar.org/paper/2a4ca461fa847e8433bab67e7bfe4620371c1f77}
\showURL{%
\tempurl}


\bibitem[\protect\citeauthoryear{Zhu and Goldberg}{Zhu and Goldberg}{2009}]%
        {Zhu2009-xp}
\bibfield{author}{\bibinfo{person}{Xiaojin Zhu} {and} \bibinfo{person}{Andrew~B
  Goldberg}.} \bibinfo{year}{2009}\natexlab{}.
\newblock \bibinfo{booktitle}{\emph{Introduction to Semi-supervised Learning}}.
\newblock \bibinfo{publisher}{Morgan \& Claypool Publishers}.
\newblock
\showISBNx{9781598295474}
\urldef\tempurl%
\url{https://doi.org/10.2200/S00196ED1V01Y200906AIM006}
\showDOI{\tempurl}


\bibitem[\protect\citeauthoryear{Zhu, Lafferty, and Ghahramani}{Zhu
  et~al\mbox{.}}{2003}]%
        {Zhu2003-dh}
\bibfield{author}{\bibinfo{person}{Xiaojin Zhu}, \bibinfo{person}{John
  Lafferty}, {and} \bibinfo{person}{Zoubin Ghahramani}.}
  \bibinfo{year}{2003}\natexlab{}.
\newblock \showarticletitle{Combining active learning and semi-supervised
  learning using gaussian fields and harmonic functions}. In
  \bibinfo{booktitle}{\emph{{ICML} 2003 workshop on the continuum from labeled
  to unlabeled data in machine learning and data mining}},
  Vol.~\bibinfo{volume}{3}.
\newblock
\urldef\tempurl%
\url{http://mlg.eng.cam.ac.uk/zoubin/papers/zglactive.pdf}
\showURL{%
\tempurl}


\end{thebibliography}

\end{document}